\documentclass[twocolumn,amssymb,nobibnotes,aps,prd,amsmath,mathtools,nofootinbib,pgfmath,pgffor,superscriptaddress]{revtex4-1}

\usepackage{graphicx}
\usepackage{amsthm}
\usepackage{amsmath}
\usepackage{amssymb}
\usepackage{booktabs}
\usepackage[dvipsnames]{xcolor}
\usepackage{hyperref}

\hypersetup{colorlinks,citecolor=blue}

\newcommand{\Nmin}{N_{\min}}

\newtheorem{proposition}{Proposition}

\begin{document}

\title{Gravitational-Wave Echoes and Greybody Cutoff from Logarithmic Throats at a Marginal-Trapping Threshold}

\author{Jahed Abedi}
\email{jahed.abedi@bimsa.cn}
\affiliation{Beijing Institute of Mathematical Sciences and Applications (BIMSA),
Huairou District, Beijing 101408, P. R. China}

\begin{abstract}
When a static, spherically symmetric configuration is brought close to a marginal-trapping threshold, it develops a long throat. As the nondegenerate minimum of $F=1-2m/r$ approaches zero, $F\simeq\Delta+\beta(r-R_M)^2$, the proper length across the throat grows as $-\ln\Delta/\sqrt\beta$, even though the local curvature stays finite. We study how axial gravitational waves propagate in an effective model of this throat, in which the lapse is parametrized as $N\propto F^{s/2}$. For $s>0$ the throat acts as a cavity behind the angular-momentum barrier, and waves come back out as echoes spaced by $T_{\rm echo}\simeq2\Lambda+\delta\propto\Delta^{-s/2}$, where $\Lambda$ is the optical length of the throat and $\delta$ does not depend on $\Delta$. We study four-dimensional, matter-supported quasi-black holes in which a regular centre caps the near-AdS$_2$ throat. These configurations provide a simplified model for investigating capped throats in horizonless microstate geometries. The echo delay in this model is controlled by the inverse redshift near the bottom of the throat, as in capped microstate geometries. This differs from thin-shell gravastars and wormholes, where the delay grows only logarithmically with the distance to the would-be horizon. Time-domain evolutions confirm the $\Delta^{-s/2}$ scaling over three decades in $\Delta$; for $s=0$ the throat becomes part of the barrier and no distinct echo appears. The barrier also imposes a greybody cutoff on the reflected spectrum, $1-\Gamma\simeq e^{-(\omega-f)/T}$, where $T$ is fixed by the curvature of the barrier and does not depend on the surface gravity of any horizon behind it. In this model, the prompt signal is the same whether the interior contains a horizon or a reflecting throat. In the horizonless case, however, the transmitted wave returns as echoes, restoring the high-frequency part of the full reflected spectrum.
\end{abstract}

\maketitle

\begin{figure*}[t]
    \centering
    \begin{tabular}{ccc}
        \begin{minipage}{0.30\textwidth}\centering\includegraphics[width=1.0\textwidth]{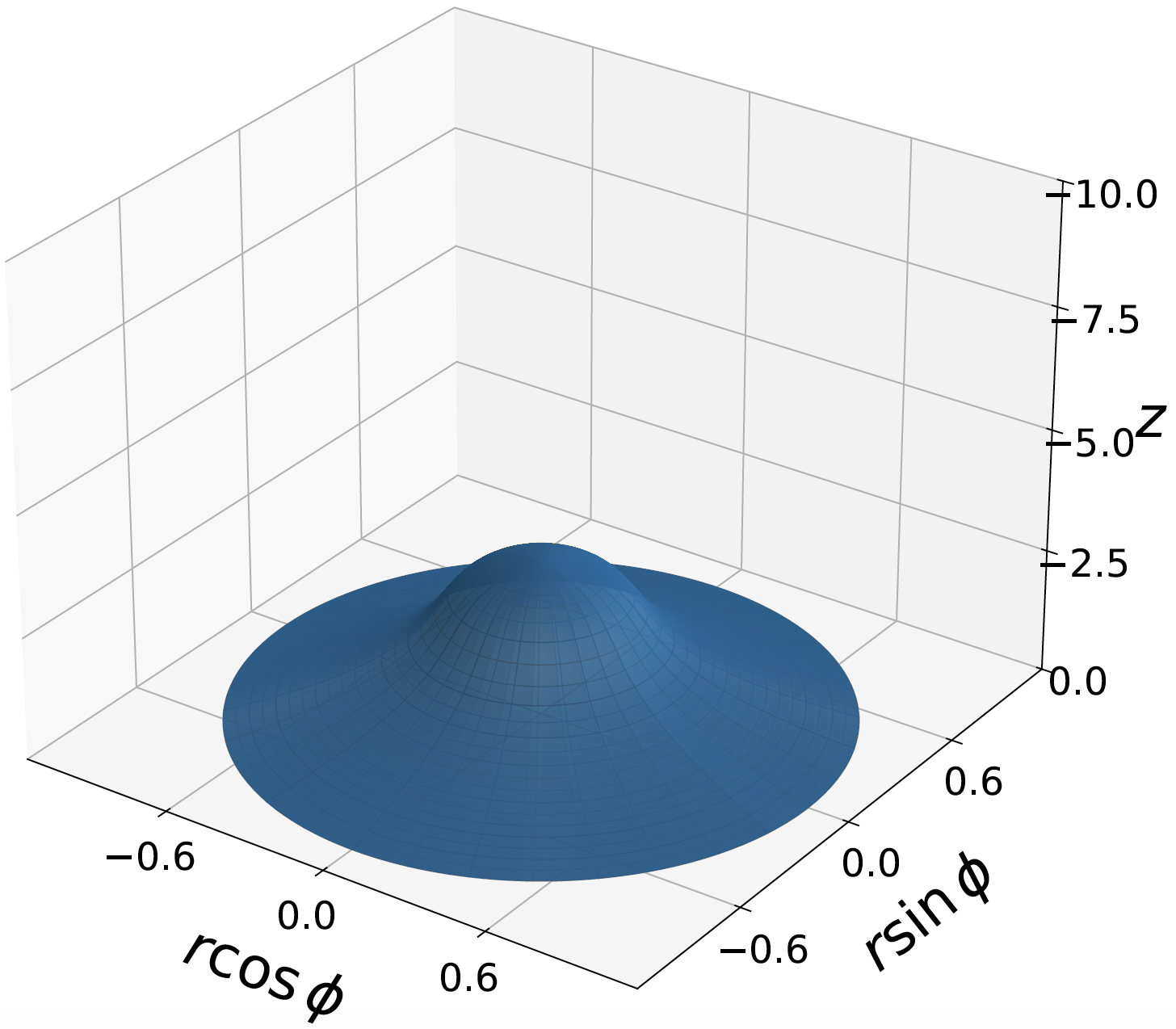}\\[0.5ex](a)\end{minipage} &
        \begin{minipage}{0.30\textwidth}\centering\includegraphics[width=1.0\textwidth]{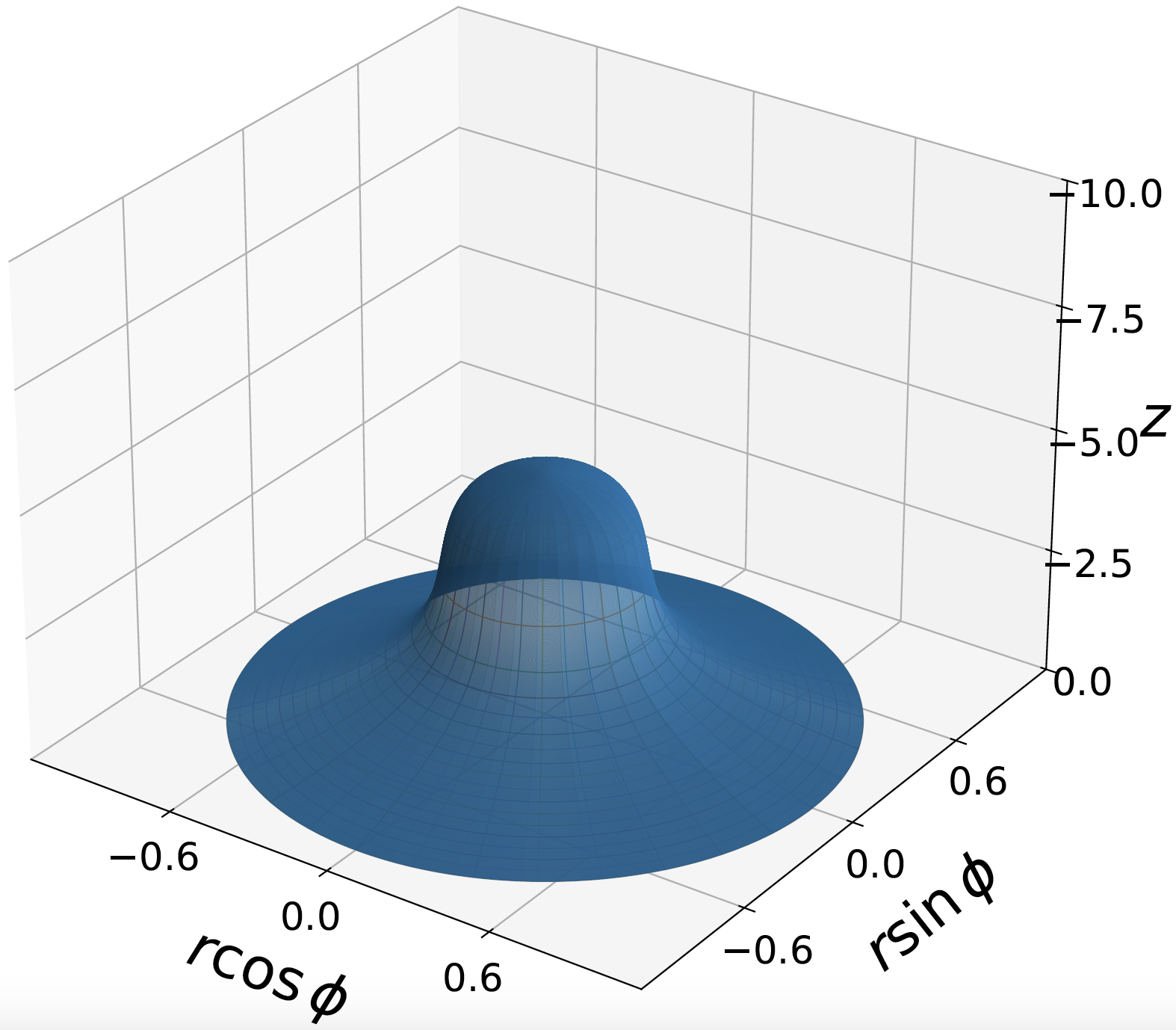}\\[0.5ex](b)\end{minipage} &
        \begin{minipage}{0.30\textwidth}\centering\includegraphics[width=1.0\textwidth]{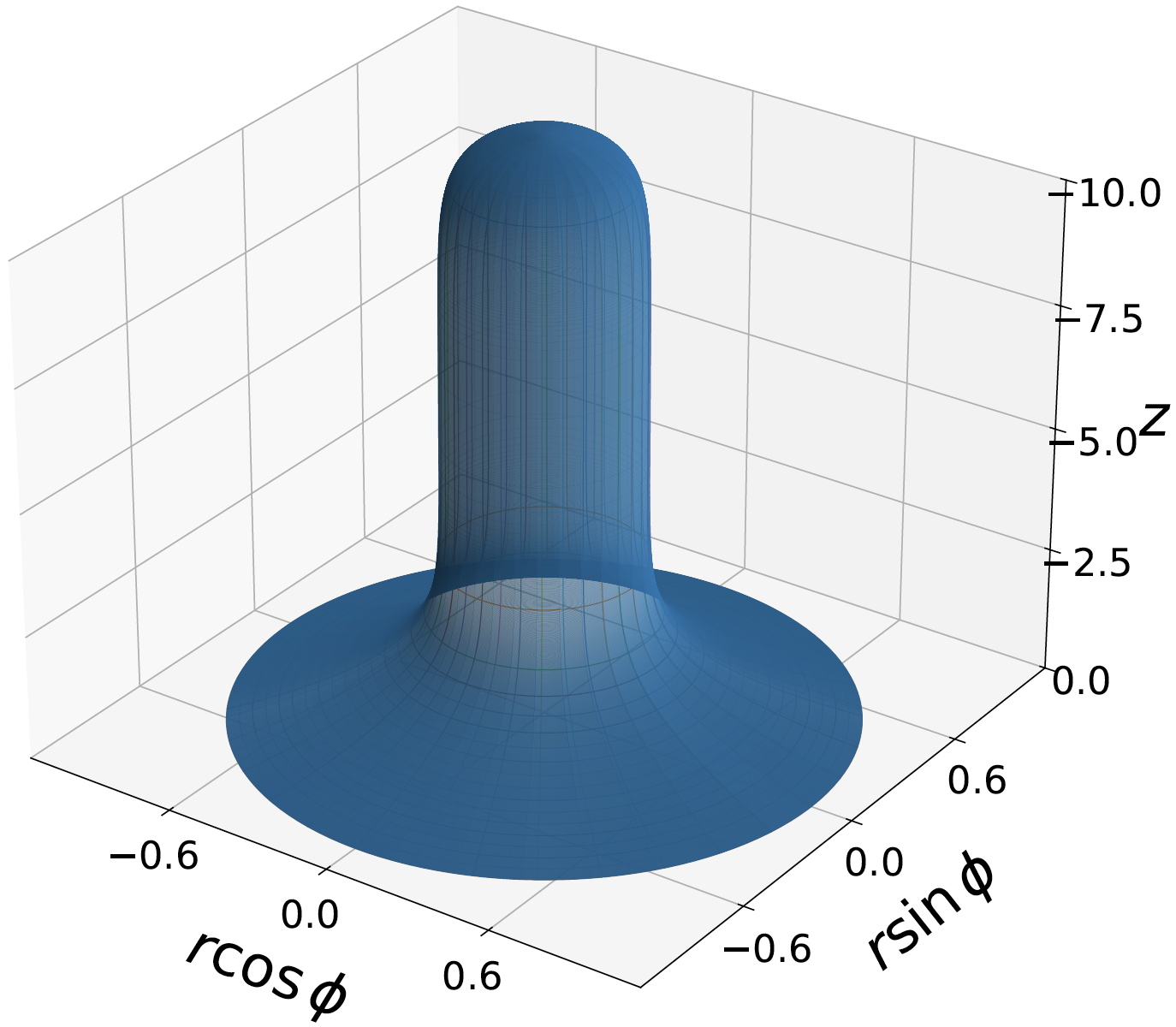}\\[0.5ex](c)\end{minipage} \\[1ex]
        \begin{minipage}{0.30\textwidth}\centering\includegraphics[width=1.0\textwidth]{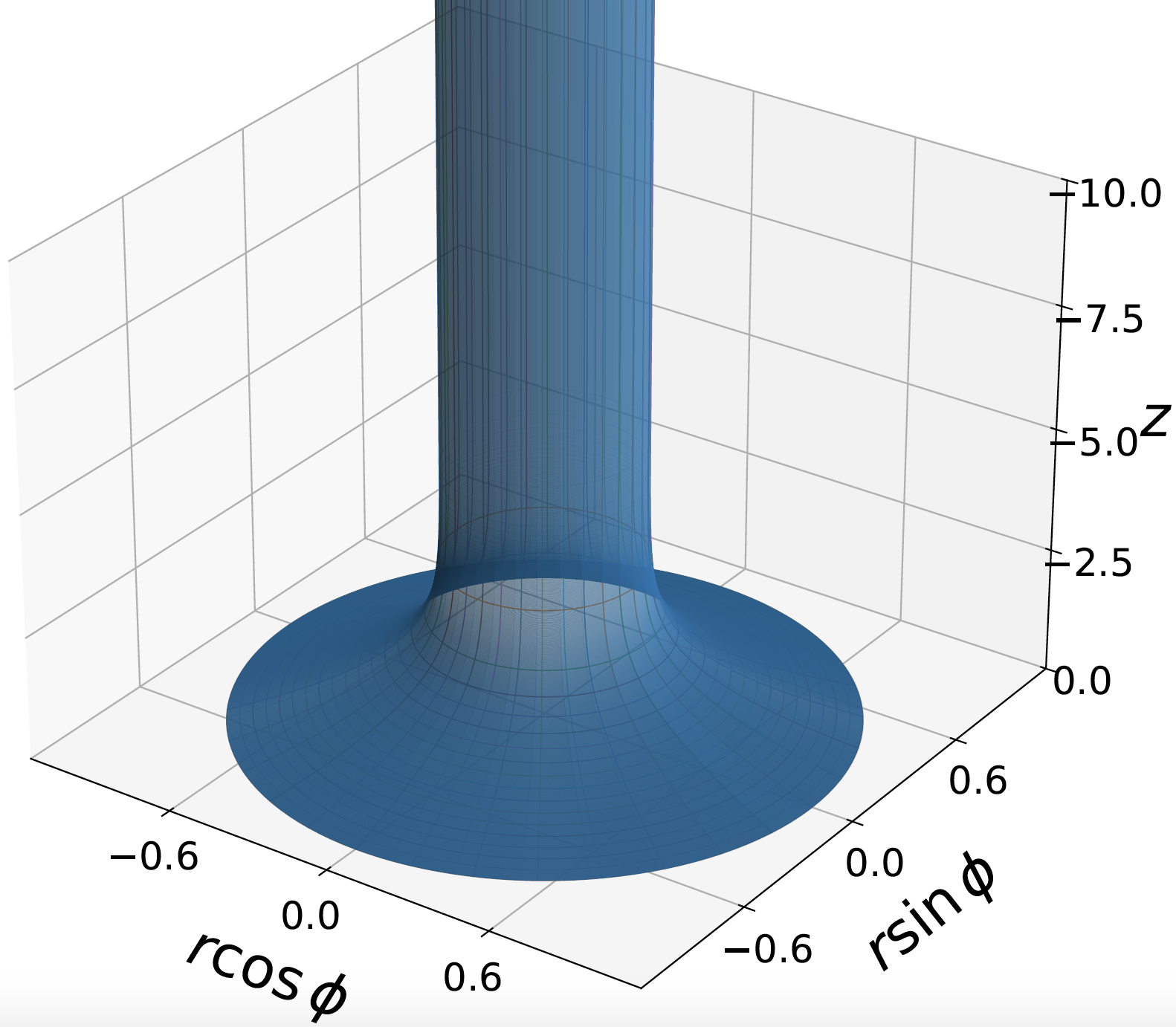}\\[0.5ex](d)\end{minipage} &
        \begin{minipage}{0.30\textwidth}\centering\includegraphics[width=1.0\textwidth]{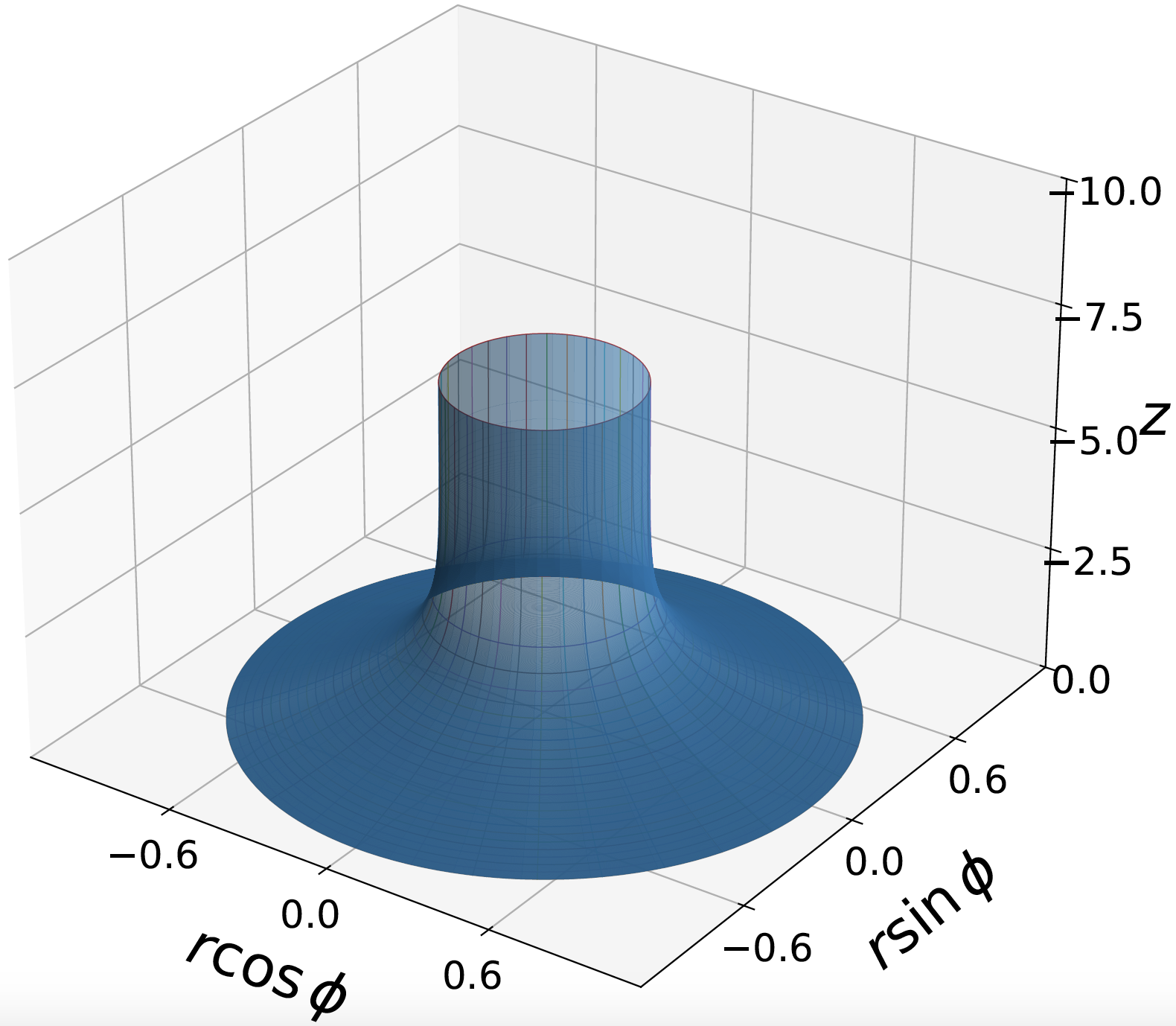}\\[0.5ex](e)\end{minipage} &
        \begin{minipage}{0.30\textwidth}\centering\includegraphics[width=1.0\textwidth]{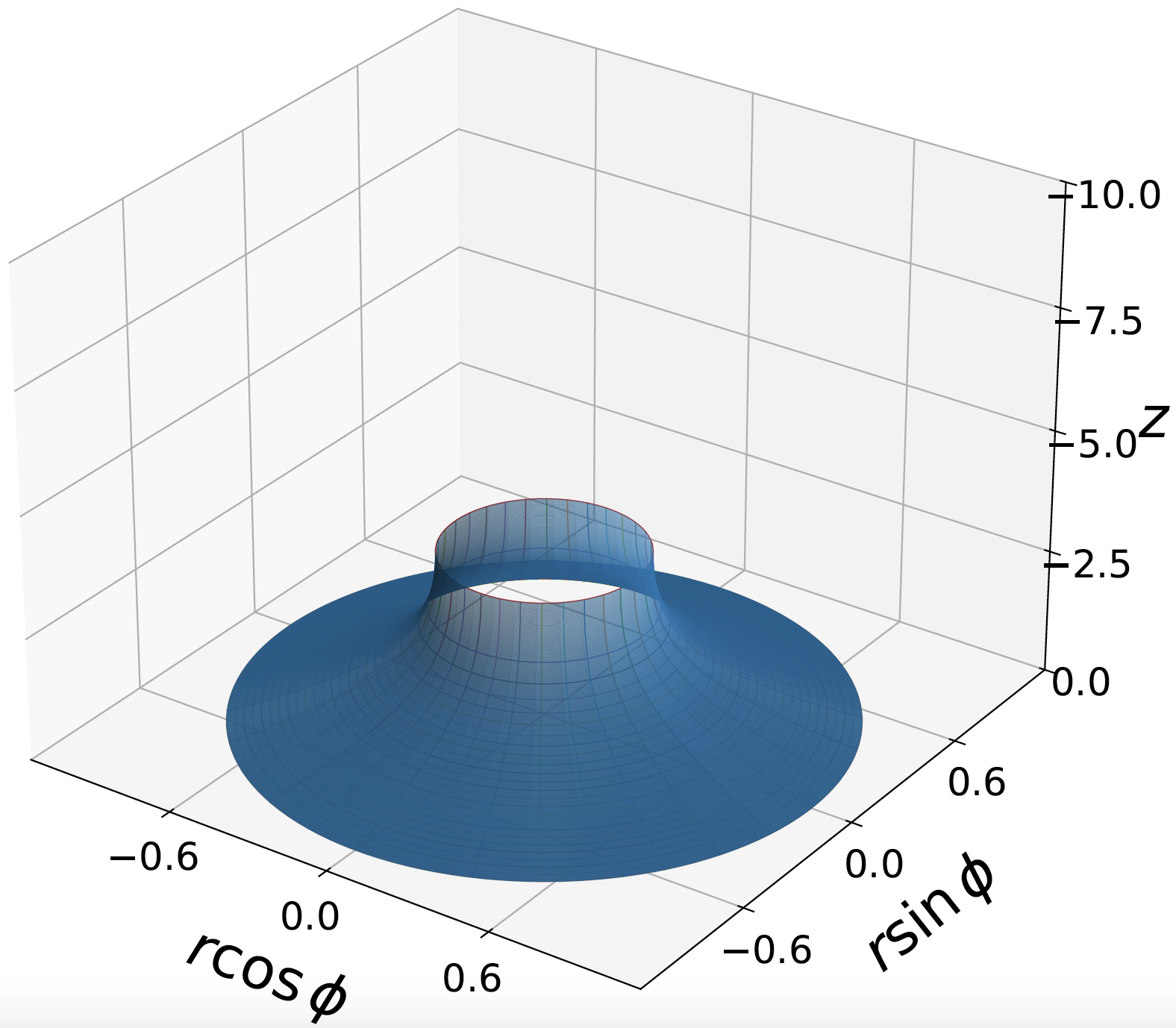}\\[0.5ex](f)\end{minipage}
    \end{tabular}
    \caption{Schematic behavior of the proper radial length and proper volume as the minimum of $F(r)$ approaches zero. The logarithmic growth occurs as $M\to M_c$. The embedding develops a throat-like shape as $F$ becomes small; this is reminiscent of spatial embedding diagrams of black-hole throats, but does not by itself imply an Einstein--Rosen bridge or wormhole \cite{Misner:1973prb,Carroll:2004st}. Panels: (a) far before the threshold ($\Delta=10^{-1.8}$); (b) approaching ($10^{-3.1}$); (c) close ($10^{-7.9}$); (d) at the onset ($10^{-11.1}$); (e) immediately after ($\Delta=-10^{-14.9}$); (f) well after ($\Delta=-10^{-3.3}$). The Burkert profile has $q=a/R=0.1$.}
    \label{fig:sixgrid}
\end{figure*}

\begin{figure*}[t]
\includegraphics[width=\textwidth]{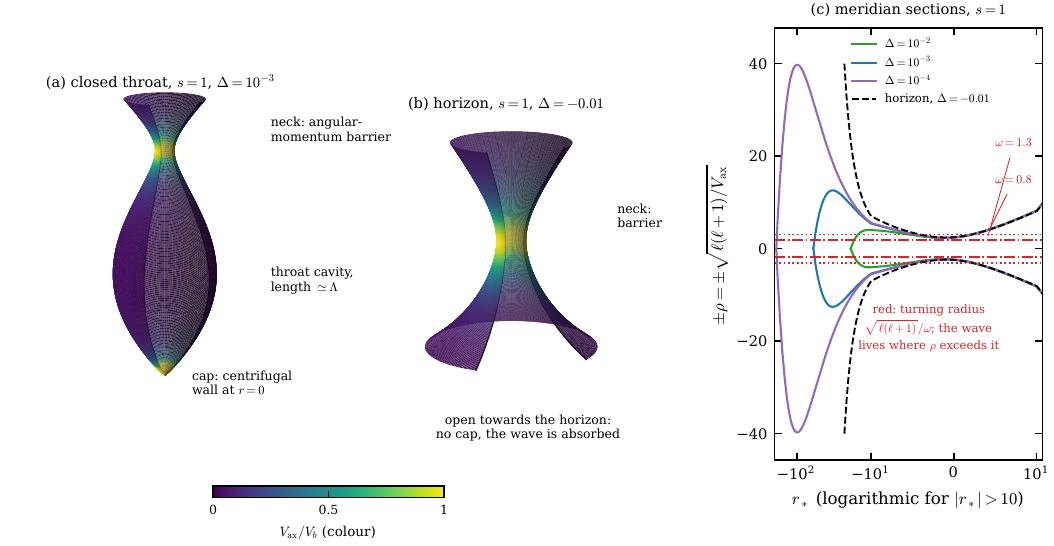}
\caption{The throat as a cavity for axial gravitational waves ($\ell=2$, units $R=1$). The surface of revolution has radius $\rho(r_*)=\sqrt{\ell(\ell+1)/V_{\rm ax}(r_*)}$, so that a wave $e^{im\phi}$ with $m^2=\ell(\ell+1)$ on it feels the potential $V_{\rm ax}$ of Eq.~\eqref{eq:Vax} as its centrifugal term (Appendix~\ref{app:waveemb}); the vertical coordinate is $r_*$ and the colour is $V_{\rm ax}/V_b$. (a) Closed throat, $s=1$, $\Delta=10^{-3}$: the centrifugal wall at $r=0$ closes the surface, the throat is a long bulb whose length along $r_*$ is the optical length $\Lambda$, and the angular-momentum barrier is the neck at its mouth. (b) Horizon, $s=1$, $\Delta=-0.01$: $V_{\rm ax}\to0$ at the horizon and the surface stays open. (c) Meridian sections $\pm\rho(r_*)$ for $s=1$ at $\Delta=10^{-2},10^{-3},10^{-4}$ and for the horizon; a wave of frequency $\omega$ propagates where $\rho>\sqrt{\ell(\ell+1)}/\omega$ (red: $\omega=0.8$ dotted, $1.3$ dash-dotted). At $\omega=0.8$ the neck is too narrow and the wave is trapped in the bulb, leaking out as echoes; at $\omega=1.3$ it passes the neck.}
\label{fig:waveemb}
\end{figure*}

\section{Introduction}

Marginally outer trapped surfaces (MOTSs) provide a quasi-local characterization of gravitational trapping through the vanishing of the outward null expansion \cite{PhysRevD.49.6467,Ashtekar:2004cn,Booth2005,GourgoulhonJaramillo2006,Ashtekar:2025wnu}. In static, spherically symmetric geometries, and for the time-symmetric slices considered here, this condition reduces to
\begin{equation}
F(r)=1-\frac{2m(r)}{r}=0,
\end{equation}
where $m(r)$ is the Misner--Sharp mass \cite{Misner:1964je}. We investigate how the geometry changes as the minimum of $F(r)$ approaches zero and what gravitational-wave signals a distant observer would measure near this threshold.

Our first result concerns the critical behavior of the spatial geometry. For a prescribed smooth mass profile, the proper radial length and the spatial volume both contain the factor $F^{-1/2}$. When a nondegenerate minimum of $F(r)$ reaches zero, these quantities, as well as the radial geodesic distance across the throat, diverge as $-\ln\Delta$, and the embedded slice develops a long throat (Fig.~\ref{fig:sixgrid}). The local curvature, by contrast, stays finite. The logarithm comes entirely from the behaviour near the minimum, so its coefficient, $1/\sqrt\beta$ for the length, is set by the curvature of the minimum and depends on the profile only through $\beta$. We work with the Burkert profile throughout. We use it as a convenient prescribed enclosed-mass family, not as a model of an astrophysical halo on the verge of collapse. This setting is different from the entropy-extremization analysis of self-gravitating radiation by Sorkin, Wald and Zhang (SWZ) \cite{Sorkin:1981wd}, who derived equilibrium configurations from the coupled Einstein-fluid equations and found that configurations arbitrarily close to their Schwarzschild radius can carry arbitrarily large entropy. Unlike the SWZ analysis, which derives equilibrium configurations from the coupled Einstein-fluid equations, we prescribe the geometry and then study its properties.

Our second result concerns gravitational-wave propagation. Because the spatial geometry does not determine the lapse, we specify a lapse family before calculating the wave dynamics. We parametrize the lapse as $N\propto F^{s/2}$ and show that it affects the throat only through three quantities: its optical length $\Lambda$, the tangential pressure and the minimum lapse. The source is then an effective anisotropic fluid without a matter Lagrangian, so its perturbations are not determined by the background and require a closure. We adopt the irrotational closure used in recent studies of axial perturbations in anisotropic matter \cite{Chakraborty2025,ZhaoPani2026} and leave the polar sector aside. Within this model, for $s>0$, the throat behaves as a cavity behind the angular-momentum barrier, and axial gravitational waves return as echoes spaced by $T_{\rm echo}\simeq2\Lambda+\delta\propto\Delta^{-s/2}$. Figure~\ref{fig:waveemb} visualizes the effective cavity by representing the axial potential through the width of a surface. In Fig.~\ref{fig:sixgrid}, the throat is represented as a bulb whose centre closes the cavity and whose narrow neck represents the angular-momentum barrier. Low-frequency waves remain trapped, whereas higher-frequency waves can cross the barrier. A horizon leaves the inner region open. Echoes have been proposed as a signature of horizonless ultracompact objects and of structure at the horizon scale \cite{CardosoFranzinPani2016,CardosoHopper2016,AbediDykaarAfshordi2017,CardosoPani2019,AbediReview2020,BertiSpectroscopy2026}. For thin-shell gravastars and wormholes, the echo delay grows logarithmically with the distance to the would-be horizon. In the present model, it instead scales as $\Delta^{-s/2}$, with the exponent determined by the lapse parameter $s$. For $s=0$ the throat becomes part of the barrier, and the packets we evolve show no distinct echo. We derive the echo-delay scaling analytically and compare it with the time-domain results.

Finally, we calculate the barrier's greybody factor to determine how the incident spectrum is divided between the prompt signal and the echoes. Above the barrier frequency the reflectivity falls off exponentially, $1-\Gamma\simeq e^{-(\omega-f)/T}$, the same form Oshita found for Kerr black holes \cite{Oshita2024}. The width $T$ is set by the curvature of the barrier, not by the surface gravity of any horizon behind it. A horizon and a throat therefore impose the same cutoff on the prompt signal. For the throat, however, the echoes return the transmitted part, so the full signal keeps its high frequencies. These wave results depend on the adopted lapse family and perturbation closure; they are not yet predictions derived from a specified matter theory.

At the critical point itself, the nondegenerate zero of $F$ lies at infinite proper distance on the static slice, and the limiting geometry develops an asymptotic cylindrical end. We therefore interpret $M=M_c$ as a geometric trapping threshold. Without a further dynamical or extension analysis, it should not be read as a complete description of regular black-hole formation.

At $M=M_c$, the zero of $F$ lies at infinite proper distance, and the static slice approaches an asymptotically cylindrical geometry. We therefore treat $M=M_c$ as a geometric trapping threshold, not as a complete model of regular black-hole formation. Establishing the latter would require a dynamical analysis or a specification of how the geometry is extended.

Table~\ref{tab:scope} lists the assumptions behind the different results: the geometric part needs only the prescribed spatial geometry, while the wave results also use the lapse family and the closure discussed below. \emph{Level~I} (Sec.~\ref{sec:geometry}) follows from the prescribed spatial geometry alone and does not involve the lapse. \emph{Level~II} (Secs.~\ref{sec:lapse} and~\ref{sec:echoes}) holds within the lapse family $N\propto F^{s/2}$; the wave results of Sec.~\ref{sec:echoes} rely in addition on the closure, the restriction to the axial sector and the wave domain used in the evolutions. The paper is organized as follows. Section~\ref{sec:lapse} introduces the lapse, the optical length and the AdS$_2$ structure of the throat. Section~\ref{sec:echoes} derives and tests the echoes and measures the greybody factor of the barrier, and Sec.~\ref{sec:conclusions} concludes. The appendices contain derivations, the curvature of the lapse family, the matter content and the construction of Fig.~\ref{fig:waveemb}.

\paragraph*{Relation to previous work.} Horizonless exotic compact objects have been proposed for many reasons, ranging from gravitational-vacuum condensate stars (gravastars) \cite{MazurMottola2004,VisserWiltshire2004,Cattoen2005} to the fuzzball and microstate-geometry programme of string theory \cite{Mathur2005,Bena2016PRL,BMMW2022}; Refs.~\cite{CardosoPani2019,Carballo-Rubio2018,Bambi2025} review them. Their generic signature is a prompt ringdown much like that of a black hole, followed by echoes of waves trapped between the angular-momentum barrier and the interior \cite{CardosoFranzinPani2016,CardosoHopper2016,AbediReview2020}. When the surface of the object sits a small distance outside a would-be nondegenerate horizon, as for thin-shell gravastars and wormholes, the echo delay grows only logarithmically with that distance \cite{CardosoHopper2016,DamourSolodukhin2007}.

The situation changes when the would-be horizon is degenerate. In the deep scaling microstate geometries, a long AdS$_2$ throat that mimics the extremal black hole ends in a smooth cap at a large redshift, controlled by a parameter of the solution \cite{Bena2016PRL,Tyukov2018}. Waves reflected from the cap come back as evenly spaced echoes, and both their period and the trapping times scale as the inverse redshift of the cap \cite{Bena2019Thermal,Bena2021Escape,Ikeda2021}. Similar cavities appear in topological stars \cite{BahHeidmann2021PRL,Heidmann2023Cavity}. Within four-dimensional general relativity, quasi-black holes, that is, static matter configurations approaching their own gravitational radius, are necessarily extremal; they develop throats of diverging proper length and infinite redshift \cite{LemosWeinberg2004,LemosZaslavskii2007} and support long-lived modes \cite{GuoQBH2022}. For $s>0$, the limit $\Delta\to0$ places this family in the quasi-black-hole class. The configurations are sourced by an uncharged anisotropic fluid, similar to the effective matter used in gravastar models \cite{Cattoen2005,Raposo2019,Brustein2022}. The proper length of its throat grows only as $\ln(1/\Delta)$, whereas the echo delay grows as the inverse of the minimum lapse, $\Delta^{-s/2}$, just as for the capped throats of microstate geometries.  We do not specify a microscopic origin for this geometry. Unlike a gravastar, it has no de Sitter core and develops a long throat near the critical limit. It provides a four-dimensional setting in which to calculate the echo delays, echo amplitudes, and greybody cutoff associated with a capped near-AdS$_2$ throat. So far, searches in gravitational-wave data have found no conclusive evidence for echoes \cite{LVK2026Remnants,BertiSpectroscopy2026}.

\begin{table}[t]
\caption{Scope of the results and the assumptions each one uses.}
\label{tab:scope}
\begin{tabular}{@{}p{0.38\columnwidth}p{0.58\columnwidth}@{}}
\toprule
Result & Assumptions\\
\midrule
Level~I: $-\ln\Delta$ growth of $\mathcal L$, $\mathcal V$ and the radial distance; finite ${}^{(3)}R$; cylindrical end (Sec.~\ref{sec:geometry}) & static, spherically symmetric, time-symmetric slice; prescribed $m(r)=Mf(r)$ whose $F$ has a nondegenerate minimum\\[0.6ex]
Level~II: optical length, AdS$_2$ throat, stresses, energy conditions, curvature (Sec.~\ref{sec:lapse}) & as above, plus the lapse family $N=N_0F^{s/2}$, with $N_0$ fixed by matching to the Schwarzschild lapse of the far field\\[0.6ex]
Wave results: echoes, ringdown, spectra, greybody factor (Sec.~\ref{sec:echoes}) & as above, plus: axial sector only; irrotational closure for the effective source (no matter Lagrangian or dynamics); Burkert profile out to $r_*=20$, mass held fixed (Schwarzschild) beyond; $\ell=2$ packets with $\omega_0=0.8$\\
\bottomrule
\end{tabular}
\end{table}

\section{Geometric setup and logarithmic stretching}
\label{sec:geometry}

\subsection{Setup}
We consider a static, spherically symmetric spacetime,
\begin{equation}
ds^2=-e^{2\Phi(r)}dt^2+\frac{dr^2}{F(r)}+r^2d\Omega^2,
\qquad
F(r)=1-\frac{2m(r)}{r},
\label{metric}
\end{equation}
where $r$ is the areal radius and $m(r)$ the Misner--Sharp mass \cite{Wald:1984rg,Misner:1964je}. The induced spatial geometry depends only on $F(r)$, so the results of this section do not involve $\Phi$. The proper radial length and spatial volume \cite{Misner:1973prb} inside $R$ are
\begin{equation}
\mathcal{L}(R)=\int_0^R\frac{dr}{\sqrt{F(r)}},
\qquad
\mathcal{V}(R)=4\pi\int_0^R\frac{r^2 dr}{\sqrt{F(r)}},
\label{spatial_observables}
\end{equation}
defined where $F>0$. If an interval with $F<0$ develops, each region with $F>0$ ends at a simple zero of $F$, where the spatial metric continues smoothly ($d^2r/d\rho^2=F'/2$ stays regular). The static ansatz does not tell us how the matter continues there, however, and distances across the throat are only meaningful on the connected slice with $\Delta>0$, so we restrict them to that case.

\begin{figure}[!tbp]
\includegraphics[width=\columnwidth]{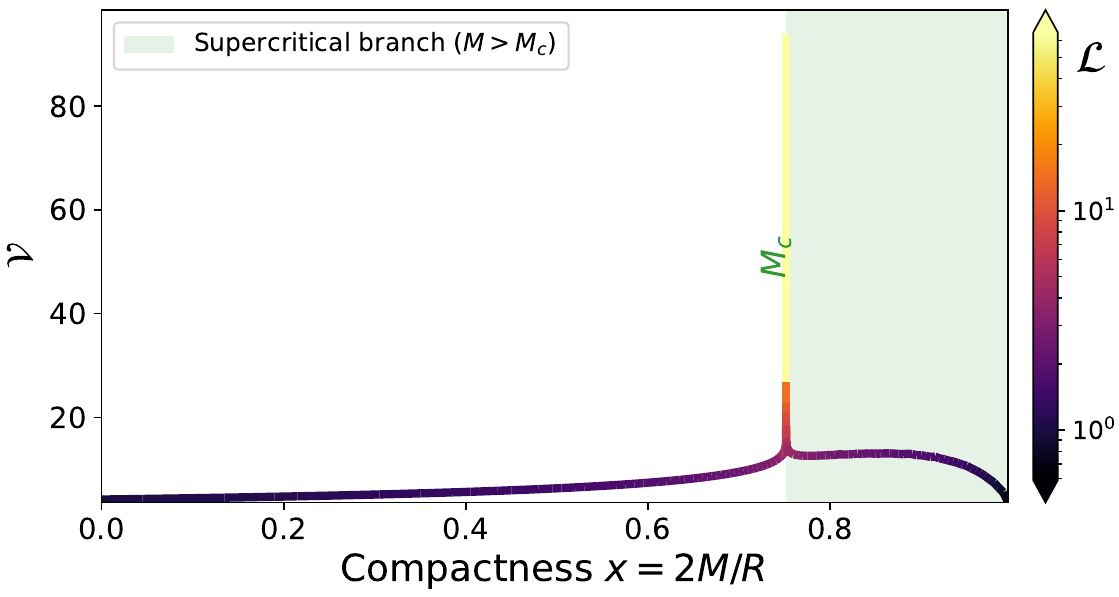}
\caption{Bulk volume $\mathcal{V}(x)$ as a function of compactness. The rapid increase near the critical region reflects the spatial stretching associated with the minimum of $F(r)$.}
\label{fig:Vol}
\end{figure}

\begin{figure}[!tbp]
    \centering
        \includegraphics[width=\columnwidth]{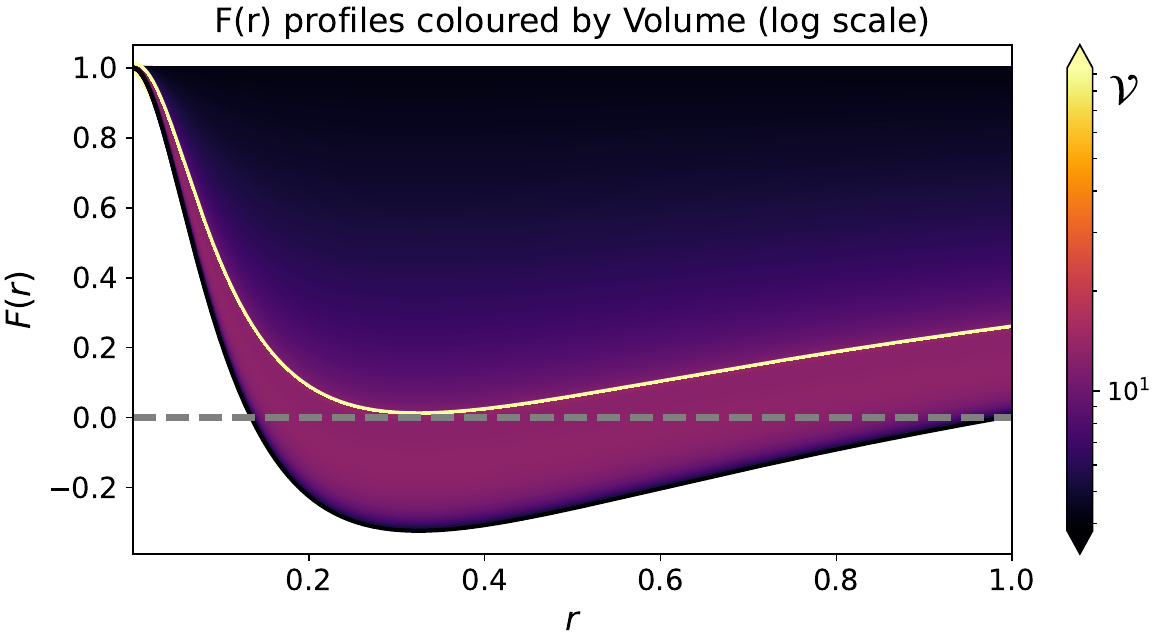}
    \caption{Profiles of $F(r)$ for different compactness values; the approach to the MOTS threshold is characterized by a minimum that approaches zero.}
    \label{fig:profiles}
\end{figure}

We consider a one-parameter family of enclosed-mass profiles $m(r)=Mf(r)$ with $f(R)=1$, labelled by the compactness $x=2M/R$, and set $R=1$. Our representative example is the Burkert profile \cite{Burkert:1995yz}, whose enclosed mass follows from the rotation curve given in Ref.~\cite{Lin:2019yux},
\begin{equation}
f(r)=
\frac{
\ln(1+\xi)+\frac12\ln(1+\xi^2)-\arctan\xi
}{
\ln(1+\xi_R)+\frac12\ln(1+\xi_R^2)-\arctan\xi_R
},
\label{burkert}
\end{equation}
with $\xi=r/a$, $\xi_R=R/a$ and $q\equiv a/R=1/\xi_R$. For $q=0.1$ the critical point is at $M_c\simeq0.3758$ (compactness $x_c\simeq0.7516$), with $R_M\simeq0.3245$ and $\beta\simeq2.81$ in units $R=1$. The choice of profile moves the critical point and changes its coefficients, but it does not change the local logarithmic scaling derived below (Figs.~\ref{fig:Vol} and~\ref{fig:profiles}). At large radius the Burkert mass grows only logarithmically, $f\propto\ln r$, so $2m/r\to0$ and $F\to1$. The profile therefore needs no exterior to be glued onto it: far from the core the geometry is asymptotically flat, just as Schwarzschild is far out. We normalize the profile at the finite radius $r=R$, and the minimum of $F$, the threshold and the throat all lie inside this radius.

Embedding the equatorial plane of a constant-$t$ slice, $dl^2=dr^2/F+r^2d\phi^2$, as a surface of revolution \cite{Misner:1973prb} gives
\begin{equation}
z(r)=\pm\int^r\frac{d\bar r}{\sqrt{\bar r/2m(\bar r)-1}} ,
\label{z_general}
\end{equation}
which produces the funnels of Fig.~\ref{fig:sixgrid}. At the threshold $z\sim-\beta^{-1/2}\ln|r-R_M|$, and the funnel becomes an infinitely long cylindrical throat.

\subsection{Critical behavior}
\label{nearcritical}
Let $M_c$ be the critical mass at which the minimum of $F(r,M)$ reaches zero, and $R_M$ the location of this minimum. This threshold is a geometric critical point of the prescribed family, and it should not be confused with the turning points and stability thresholds of self-gravitating radiation equilibria \cite{Sorkin:1981wd,Chavanis:2001ah,Chavanis:2007kn}. It is a degenerate critical point of the spatial geometry rather than a MOTS at finite radius. For $M<M_c$ the slice contains no MOTS; for $M>M_c$ the simple zeros of $F$ are MOTSs (minimal spheres of the time-symmetric slice) at finite distance; and at $M=M_c$ the limiting sphere sits at infinite proper distance. At the threshold
\begin{equation}
\begin{gathered}
F(R_M,M_c)=0,\qquad \partial_rF|_{R_M}=0,\\
\beta\equiv\tfrac12\partial_r^2F|_{R_M,M_c}>0 .
\end{gathered}
\label{critical_conditions}
\end{equation}
Define $\alpha\equiv-\partial_MF(R_M,M)|_{M_c}>0$. For $M<M_c$,
\begin{equation}
F(r,M)\simeq\alpha(M_c-M)+\beta(r-R_M)^2+\cdots .
\label{Fexpansion}
\end{equation}

\paragraph*{The mass-scaling family.}
For $m=Mf(r)$ it is convenient to write $g=f/r$, so that $F=1-2Mg(r)$. The minimum of $F$ is then the maximum of $g$, which means that $R_M$ does not move as $M$ changes, $M_c=1/(2g_{\max})$, and $F(R_M)=\Delta$ exactly. Hence $\alpha M_c=1$ and, with $\zeta\equiv r-R_M$,
\begin{equation}
F=1-(1-\Delta)\frac{g}{g_{\max}}=\Delta+(1-\Delta)\,\beta\,\zeta^2+O(\zeta^3),
\label{eq:Fclean}
\end{equation}
where
\begin{equation}
\beta=-\frac{g''(R_M)}{2g_{\max}}
\label{eq:beta}
\end{equation}
is exact for the mass-scaling family when $\beta$ is defined at $M_c$ as in Eq.~\eqref{critical_conditions}. In what follows we write $F\simeq\Delta+\beta\zeta^2$. The factor $(1-\Delta)$ changes $\beta$ only at $O(\Delta)$, so it enters the expansions below only through terms of order $\Delta\ln\Delta$.

At the threshold, with $\rho=\beta^{-1/2}\ln|\zeta|$, the line element becomes $dl^2=d\rho^2+R_M^2d\Omega^2$. The critical sphere is pushed to infinite proper distance, and the geometry develops asymptotically cylindrical ends of finite areal radius $R_M$. The two sides of the developing throat drift infinitely far apart, but this alone does not join them into an Einstein--Rosen bridge. The spatial throat looks like that of extremal Reissner--Nordstr\"om, although we do not claim that the spacetime is extremal Reissner--Nordstr\"om. The logarithmic divergence arises from approaching a zero minimum of a spherical $F(r,M)$ from a horizonless configuration, rather than from a horizon geometry that is already extremal \cite{Carroll:2009maa,Harlow:2022ich}. The cylindrical end is also reminiscent of the quasi-black-hole limit \cite{LemosZaslavskii2007}.

After the substitution $u=(r-R_M)\sqrt{\beta/\alpha|M_c-M|}$, the integration range grows as $|M_c-M|^{-1/2}$. Introducing a cutoff $X$ in $\zeta$, beyond which the quadratic form no longer applies, the two-sided integral is
\begin{equation}
\int_{-X}^{X}\frac{d\zeta}{\sqrt{\Delta+\beta \zeta^2}}
=\frac{1}{\sqrt\beta}\Bigl[\ln\frac1\Delta+2\ln\bigl(2X\sqrt\beta\bigr)\Bigr]+O(\Delta),
\label{eq:twosided}
\end{equation}
which gives
\begin{equation}
\boxed{
\mathcal{L}(M)=-\frac{1}{\sqrt{\beta}}\ln|\Delta|+\mathcal{L}_0+\cdots ,
}
\label{Lnear}
\end{equation}
with $\Delta\equiv(M_c-M)/M_c$. The coefficient does not depend on $\alpha$, because $\ln(\alpha M_c\Delta)$ differs from $\ln\Delta$ only by a constant, which is absorbed into $\mathcal{L}_0$. In the region that dominates the integral $r$ hardly changes, so we may replace $r^2$ by $R_M^2$ and obtain
\begin{equation}
\boxed{
\mathcal{V}(M)=-\frac{4\pi R_M^2}{\sqrt{\beta}}\ln|\Delta|+\mathcal{V}_0+\cdots .
}
\label{Vnear}
\end{equation}

\paragraph*{Supercritical side.}
For $M>M_c$, $F\simeq\beta(\zeta^2-\epsilon^2)$ with $\epsilon^2=\alpha(M-M_c)/\beta$ ($=|\Delta|/\beta$ for the mass-scaling family), and $F<0$ for $r_-<r<r_+$, $r_\pm=R_M\pm\epsilon$. We define
\begin{align}
\mathcal{L}_>(M)&=\int_0^{r_-}\frac{dr}{\sqrt F}+\int_{r_+}^{R}\frac{dr}{\sqrt F},\\
\mathcal{V}_>(M)&=4\pi\Bigl[\int_0^{r_-}+\int_{r_+}^{R}\Bigr]\frac{r^2\,dr}{\sqrt F}.
\end{align}
Each branch contributes $\beta^{-1/2}\,\mathrm{arccosh}(X/\epsilon)\simeq(2\sqrt\beta)^{-1}\ln(1/|\Delta|)$, half the coefficient of the subcritical two-sided integral, so together the two branches reproduce Eqs.~\eqref{Lnear} and~\eqref{Vnear} with $|\Delta|$ in place of $\Delta$. Note that for $M>M_c$ these are the lengths of two disconnected regions, not a distance between points.

The logarithm is a local result: it follows from the nondegenerate quadratic minimum alone and does not depend on the global form of the profile. If the quadratic coefficient vanished and the first nonzero term were $O[(r-R_M)^{2n}]$ with $n\ge2$, the divergence would instead be a power law, $\mathcal{L}_{\rm sing},\mathcal{V}_{\rm sing}\sim|\Delta|^{-(n-1)/2n}$. Reaching that case requires tuning more than one parameter, so it is not generic for a one-parameter family. Not every profile has an interior minimum of $F$, though. The condition $F'(R_M)=0$ is equivalent to $\rho_E(R_M)=\bar\rho(R_M)/3$, where $\bar\rho$ is the mean enclosed density, so the local density must drop below one third of the mean enclosed density somewhere. A uniform-density star never satisfies this: its $F$ is monotonic, the minimum sits on the boundary where $F'\ne0$, and no logarithm appears.

\begin{figure}[!tbp]
\includegraphics[width=\columnwidth]{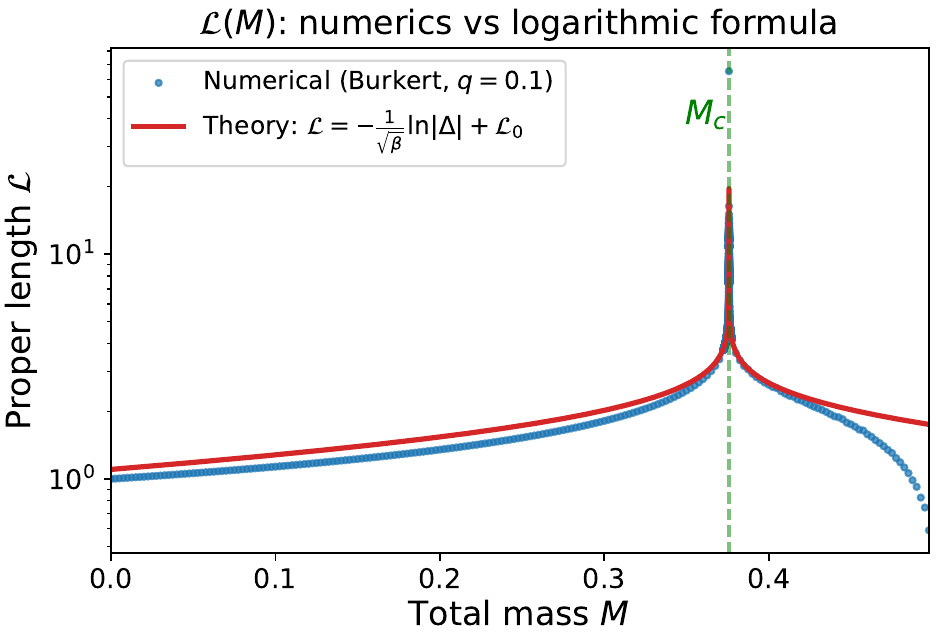}
\caption{Numerical (points) and analytical (line) proper radial length $\mathcal{L}(M)$ for the Burkert profile. The analytical curve is the logarithmic prediction of Eq.~\eqref{Lnear}.}
\label{fig:L_vs_M}
\end{figure}

\paragraph*{Numerical check.}
For the mass-scaling family the slope of $\mathcal{L}$ against $\ln(1/\Delta)$ contains no free parameter: $d\mathcal{L}/d\ln(1/\Delta)\to1/\sqrt\beta$. For the Burkert profile with $q=0.1$, the curvature of $g$ at its maximum gives $\beta=2.8065$ and hence $1/\sqrt\beta=0.59692$, while a linear fit of $\mathcal{L}$ against $\ln(1/\Delta)$ over seven values $10^{-10}\le\Delta\le10^{-4}$ gives $0.59696\pm0.00002$. The difference $\mathcal{L}-\beta^{-1/2}\ln(1/\Delta)$ settles to a constant, $\mathcal{L}_0\simeq0.9500$ (it is $0.9494$, $0.9500$, $0.9500,\dots$ for $\Delta=10^{-4},10^{-5},10^{-6},\dots$). The same fit for $\mathcal{V}$ gives a slope ratio $\mathcal{V}_{\rm sing}/\mathcal{L}_{\rm sing}=1.3235$, compared with the local prediction $4\pi R_M^2=1.3229$; the difference of $4\times10^{-4}$ is within the $O(\Delta\ln\Delta)$ corrections expected in the fit window. Fig.~\ref{fig:L_vs_M} compares $\mathcal{L}(M)$ with Eq.~\eqref{Lnear} over many decades of $\Delta$.

\paragraph*{Geodesic length across the throat.}
For $\Delta>0$ the same divergence shows up in the length of geodesics that connect fixed points on opposite sides of the throat. Consider a geodesic on the constant-$t$ slice from $r_1<R_M$ to $r_2>R_M$ with conserved angular momentum $J<\min(r_1,r_2)$, so that $r$ increases monotonically along it. It has $dl=dr/\sqrt{F(1-J^2/r^2)}$, and its length is
\begin{equation}
\boxed{
\mathcal{D}(r_1,r_2;J)=-\frac{\ln\Delta}{\sqrt{\beta}\sqrt{1-J^2/R_M^2}}+\mathcal{D}_0+\cdots .
}
\label{Ddiv}
\end{equation}
At fixed $J\ne0$, however, the angle swept by the geodesic also grows, as $J\ln(1/\Delta)/(R_M^2\sqrt\beta\sqrt{1-J^2/R_M^2})$, so such geodesics wind around the throat and do not keep joining fixed endpoints; for fixed endpoints $J$ must decrease as $\Delta\to0$. For endpoints on the same radial line the minimizing geodesic is the radial one, $J=0$, because $dl\ge dr/\sqrt F$; its length is the intrinsic distance and carries the coefficient $1/\sqrt\beta$.

\subsection{Curvature, density and the cylindrical end}
\label{sec:curvature}
\paragraph*{Local curvature stays finite.}
In the proper radial coordinate, $d\ell^2=d\rho^2+r(\rho)^2d\Omega^2$, the three-dimensional Ricci scalar is
\begin{equation}
{}^{(3)}R=\frac{2}{r^2}\Bigl[1-\Bigl(\frac{dr}{d\rho}\Bigr)^2\Bigr]-\frac4r\frac{d^2r}{d\rho^2},
\end{equation}
and $dr/d\rho=\sqrt F$, $d^2r/d\rho^2=F'/2$. At the critical point $F=F'=0$, so
\begin{equation}
\boxed{{}^{(3)}R\to\frac{2}{R_M^2}.}
\label{eq:R3-critical}
\end{equation}
The critical behavior is therefore not a local curvature singularity. Instead, the slice degenerates globally into an infinitely long cylindrical end.

\paragraph*{Integrated curvature and density.}
In the throat $dV\simeq4\pi R_M^2d\rho$, so
\begin{equation}
\int{}^{(3)}R\,dV\simeq8\pi\mathcal{L}\sim-\frac{8\pi}{\sqrt\beta}\ln|\Delta| ,
\label{eq:integrated-R}
\end{equation}
which diverges logarithmically even though the local curvature is finite. On a time-symmetric slice ${}^{(3)}R=16\pi\rho_E$, so $\rho_E\to1/(8\pi R_M^2)$ and
\begin{equation}
\int\rho_E\,dV\simeq\frac{\mathcal{L}}{2}\sim-\frac{1}{2\sqrt\beta}\ln|\Delta| .
\label{eq:matter-integral}
\end{equation}
The same quantity appears as the singular part of the binding energy $E_B(R)=\int_0^R m'(r)F^{-1/2}dr-M$ \cite{Misner:1973prb}. That singular part is $E_B^{\rm sing}=-m'(R_M)\beta^{-1/2}\ln|\Delta|$, and at the threshold the conditions $F'=0$ and $2m=R_M$ give $m'(R_M)=m/R_M=\tfrac12$, so $E_B^{\rm sing}=\mathcal{L}/2$, in agreement with Eq.~\eqref{eq:matter-integral}. The proper energy integral thus diverges while $M=m(R)$ stays finite, which agrees qualitatively with the spatial-stretching mechanism of SWZ \cite{Sorkin:1981wd}. Since this argument rests on the time-symmetric Hamiltonian constraint, it should be read as a consistency check on the geometry rather than as a physical prediction.

\section{The lapse and the optical length}
\label{sec:lapse}

\subsection{Matter content for an arbitrary lapse}
Primes denote $d/dr$, and $\partial_\rho=\sqrt F\,\partial_r$ is the derivative with respect to the proper radial coordinate. The constraint equations determine $F$ from the density, $8\pi\rho_E=(1-F-rF')/r^2$, but say nothing about $\Phi$. Whatever $\Phi$ is chosen, the remaining Einstein equations and the conservation law then define the stresses,
\begin{equation}
\begin{gathered}
8\pi p_r=\frac{F(1+2r\Phi')-1}{r^2},\\
p_t=p_r+\frac r2\bigl[p_r'+(\rho_E+p_r)\Phi'\bigr].
\end{gathered}
\label{eq:stresses}
\end{equation}
At the critical point $\rho_E=1/(8\pi R_M^2)$. For lapses that are regular in the proper radial coordinate, that is, with $\partial_\rho\Phi$ bounded so that $F\Phi'\to0$, one finds $p_r=-\rho_E$. In the cylindrical region the metric is a product $\mathbb{R}^2_{(t,\rho)}\times S^2_{R_M}$, and with $N\equiv e^\Phi$
\begin{equation}
G^t{}_t=G^\rho{}_\rho=-\frac1{R_M^2},\qquad
G^\theta{}_\theta=\frac{N_{,\rho\rho}}{N}
\;\Rightarrow\;
p_t=\frac{N_{,\rho\rho}}{8\pi N}.
\label{eq:pt}
\end{equation}
The lapse therefore enters the matter content only through $p_t$, and the dominant energy condition limits how fast it can grow, $|N_{,\rho\rho}/N|\le R_M^{-2}$. Physically, the lapse sets the profile of the tangential pressure, and with it $\rho_E+p_t$ and the dominance conditions. For $N\propto e^{\sqrt\beta\rho}$ one finds $p_t=\beta/(8\pi)$, which equals $\rho_E$ only if $\beta R_M^2=1$; this is the case for extremal Reissner--Nordstr\"om, whereas the Burkert profile has $\beta R_M^2=0.296$.

An isotropic perfect fluid cannot reach the threshold, because its pressure diverges before the throat forms (Appendix~\ref{app:matter}). The lapse family introduced below should therefore be read as a parametrization of possible geometries, not as a prediction for a specified kind of matter.

\subsection{Optical length and a one-parameter family}
We define the optical length $\Lambda=\int d\rho/N=\int dr/(N\sqrt F)$, which is the light-crossing time measured in $t$, and the tortoise coordinate $dr_*=dr/(N\sqrt F)$. Writing $\psi=\chi(r_*)Y_{\ell m}e^{-i\omega t}/r$, the scalar wave equation becomes
\begin{equation}
\begin{gathered}
\chi_{,r_*r_*}+(\omega^2-V_\ell)\chi=0,\\
V_\ell=N^2\Bigl[\frac{\ell(\ell+1)}{r^2}+\frac{F'/2+\Phi'F}{r}\Bigr].
\end{gathered}
\label{eq:master}
\end{equation}
Since $r\simeq R_M$ in the throat, the last term is only a small correction there. For $\ell=0$ the equation reduces, to leading order in the throat, to the free one-dimensional wave equation on an interval of length $\Lambda$ (the residual potential $V_0=N^2(F'/2+\Phi'F)/r$ is small there but not zero); for $\ell\ge1$ the potential is $N^2\ell(\ell+1)/R_M^2$. The lapse therefore enters only through $\Lambda$ and the profile of $N$.

Rather than fixing the lapse, we parametrize it as
\begin{equation}
N=N_0F^{s/2},\qquad s\ge0,
\label{eq:family}
\end{equation}
which gives
\begin{equation}
\Lambda\simeq
\begin{cases}
\dfrac{\ln(1/\Delta)}{N_0\sqrt\beta}, & s=0,\\[2mm]
\dfrac{c_s\,\Delta^{-s/2}}{N_0\sqrt\beta},\ \ c_s=\dfrac{\sqrt\pi\,\Gamma(s/2)}{\Gamma(\frac{1+s}{2})}, & s>0,
\end{cases}
\label{eq:Lambda}
\end{equation}
with $c_1=\pi$. The level spacing $\pi/\Lambda$ of an idealized reflecting throat therefore shrinks as $1/\ln(1/\Delta)$ for $s=0$ and as $\Delta^{s/2}$ for $s>0$.

The constant $N_0$ sets the normalization of $t$. We choose it so that $N$ matches the Schwarzschild lapse $\sqrt F$ at the outer radius $r_c$ of the wave domain (defined in Sec.~\ref{sec:echoes}; beyond $r_c$ the mass is held fixed, so the far field is exactly Schwarzschild). This gives $N_0=F(r_c)^{(1-s)/2}$, so that $N_0=1$ for $s=1$, and $N_0=0.939$, $0.967$, $1$ and $1.084$ for $s=0,\tfrac12,1,2$. These values change with $\Delta$ only at $O(\Delta)$, and all values of $\Lambda$ quoted below use them. At $\Delta=0$ and $s>0$ the lapse vanishes on the critical surface, which becomes a degenerate Killing horizon. For $s=0$, on the other hand, the lapse stays finite, $N\to N_0$, and the critical geometry is an infinitely long horizonless cylinder rather than a horizon.

As $\Delta\to0$ with $s>0$, $N\propto e^{s\sqrt\beta\rho}$, and the near-horizon geometry is AdS$_2$ of radius $\ell_2=1/(s\sqrt\beta)$, with $p_t=1/(8\pi\ell_2^2)$. For $s=1$ (the gauge $e^{2\Phi}=F$) the whole near-throat region is, to leading order in the quadratic expansion of $F$, global AdS$_2$. Bertotti--Robinson is the special case $\ell_2=R_M$, i.e.\ $s=1/(\sqrt\beta R_M)$; because $p_r=-\rho_E$ at the throat, it then has $p_t=\rho_E=-p_r$, which is the stress of a radial electric field. The dominant energy condition, $|p_t|\le\rho_E$, becomes $s^2\beta R_M^2\le1$.

\paragraph*{The case $s=1$: global AdS$_2$.}
Within the quadratic approximation $F\simeq\Delta+\beta\zeta^2$, with $\zeta=\sqrt{\Delta/\beta}\sinh\sigma$ one has $F=\Delta\cosh^2\sigma$, $d\rho=d\sigma/\sqrt\beta$ and $N=N_0\sqrt\Delta\cosh\sigma$, so the throat is global AdS$_2$ with time coordinate $\tau=N_0\sqrt{\beta\Delta}\,t$. Its normal modes are \cite{Higuchi2022}
\begin{equation}
\begin{gathered}
\omega_{n\ell}=N_0\sqrt{\beta\Delta}\,(n+h_\ell),\\
h_\ell=\tfrac12+\sqrt{\tfrac14+\ell(\ell+1)\ell_2^2/R_M^2},
\end{gathered}
\end{equation}
For $\ell\ge1$ we have $h_\ell>3/2$, and square-integrability alone selects these frequencies. For $\ell=0$ the boundary condition at the ends of the throat is a genuine choice \cite{Higuchi2022}, and which branch the regular centre and the exterior select depends on how the throat is matched to them; we do not determine it here. These are the frequencies of the test scalar of Eq.~\eqref{eq:master}. For axial gravitational waves the throat potential tends to $N^2(\ell-1)(\ell+2)/R_M^2$ (Sec.~\ref{sec:echoes}), so $\ell(\ell+1)$ in $h_\ell$ is replaced by $(\ell-1)(\ell+2)$ while the spacing stays the same. All of these are asymptotic frequencies of the near-throat region; at finite $\Delta$ they are corrected by the non-quadratic terms of $F$ and by the region outside the throat, as we now quantify.

\paragraph*{Numerical check of the spectrum.}
We solved Eq.~\eqref{eq:master} in the $e^{2\Phi}=F$ spacetime ($s=1$, $N_0=1$, with a regular centre and a Dirichlet box far outside the throat) and identified the throat-localized states by the weight of their eigenvectors. At $\Delta=10^{-6}$ the frequencies $\omega_{n\ell}/\sqrt{\beta\Delta}$ are $3.1494,4.1494,5.1493,\dots$ for $\ell=1$ and $5.034,6.034,7.034,\dots$ for $\ell=2$, compared with $n+h_\ell$ for $h_1=3.1494$ and $h_2=5.0341$. The deviations shrink as $\Delta\to0$; for $\ell=1$, $n=0$ the values are $3.1417$, $3.1486$ and $3.1493$ at $\Delta=10^{-3},10^{-4},10^{-5}$. This confirms the AdS$_2$ spectrum and the formula for $h_\ell$ at $s=1$ for $\ell\ge1$. The $\ell=0$ modes, which see no centrifugal barrier, are sensitive to how the throat is matched to the centre and to the exterior, and we leave them aside.

\paragraph*{Stresses for $s=1$.}
For $s=1$ ($N=\sqrt F$, $N_0=1$), Eq.~\eqref{eq:stresses} gives at every $r$
\begin{equation}
p_r=-\rho_E,\qquad p_t=-\rho_E-\tfrac r2\rho_E' .
\end{equation}
This is the familiar $g_{tt}g_{rr}=-1$ class used for regular-core geometries, and it includes extremal Reissner--Nordstr\"om as the case $\rho_E\propto r^{-4}$, $p_t=+\rho_E$. For the Burkert density the null and weak energy conditions hold at every radius ($\rho_E+p_r=0$ and $\rho_E+p_t=-r\rho_E'/2\ge0$). The dominant condition also holds, because the logarithmic slope $-d\ln\rho_E/d\ln r$ lies in $[0,3)$, below $4$, while the strong condition is violated in the core, as it is for generic regular cores. The Kretschmann scalar $K=(F'')^2+4(F')^2/r^2+4(1-F)^2/r^4$ (Appendix~\ref{app:curv}) is finite at $R_M$ and at the Burkert core. The source is an effective anisotropic one, as in anisotropic ultracompact stars \cite{Raposo2019}. With $p_r=-\rho_E$ it shares the radial equation of state of the frozen-star model \cite{Brustein2022,Brustein2024}, which, however, has $p_t=0$ in its bulk (see also Sec.~3.3.7 of Ref.~\cite{BertiSpectroscopy2026}). Since no matter Lagrangian and no dynamics are specified, its perturbations are not determined by the background; Sec.~\ref{sec:echoes} treats axial perturbations under a stated closure and leaves the polar sector aside. The Burkert density therefore determines the geometry, but it does not by itself specify a complete matter model.

\paragraph*{Energy conditions for general $s$.}
For $s\ne1$ the radial combination $8\pi(\rho_E+p_r)=(s-1)F'/r$ changes sign at the throat, so the null energy condition fails on one side of it. Of all the members of the family, only $s=1$ satisfies the null, weak and dominant conditions everywhere (Appendix~\ref{app:matter}).

\paragraph*{Curvature.}
For $\Delta>0$ the lapse is positive and smooth, the spacetime has no horizon, and for every $s$ the curvature components in the static orthonormal frame are bounded uniformly in $\Delta$. In particular, the Kretschmann scalar stays finite: as $\Delta\to0$ it tends to $4s^2\beta^2+4/R_M^4$ at the minimum, and to $4s^4\beta^2+4/R_M^4$ on the AdS$_2$ plateau $\sqrt{\Delta/\beta}\ll|\zeta|\ll R_M$. At $\Delta=0$ and $s>0$ the surface $\zeta=0$ becomes a degenerate Killing horizon. There the curvature measured by an infalling observer diverges unless $s=1$ or $0<s\le\tfrac12$, even though every scalar invariant, the Kretschmann scalar included, remains finite (Proposition~\ref{prop:curv}, Appendix~\ref{app:curv}).

\section{Gravitational-wave echoes}
\label{sec:echoes}

A packet of gravitational waves arriving from far away is first scattered by the angular-momentum barrier, which produces the prompt ringdown \cite{KokkotasSchmidt1999,Nollert1999,BertiCardosoStarinets2009,KonoplyaZhidenko2011,PriceNampalliwarKhanna2016,BertiSpectroscopy2026}. For $\Delta>0$ and $s>0$, the part transmitted through the barrier enters the throat, which acts as a cavity, and comes back out after a delay as an echo (Fig.~\ref{fig:waveemb}). This is the same mechanism that has been proposed for horizonless ultracompact objects and for structure at the horizon scale \cite{CardosoFranzinPani2016,CardosoHopper2016,AbediDykaarAfshordi2017,PriceKhanna2017,MarkZimmerman2017,WangAfshordi2018}, reviewed in Refs.~\cite{CardosoPani2019,AbediReview2020,BertiSpectroscopy2026}. Echoes have also been studied for ultracompact and anisotropic stars \cite{Kokkotas1995,FerrariKokkotas2000,PaniFerrari2018,Raposo2019}, whose matter content is close to that of the throat considered here. Searches in LIGO--Virgo--KAGRA data have reported tentative evidence as well as null results \cite{AbediDykaarAfshordi2017,ConklinHoldomRen2018,Westerweck2018,LoLiWeinstein2019,Uchikata2019,LVK2021GWTC2TGR,AbediLongoMicchiAfshordi2023,Abedi2025echoes}, and so far there is no conclusive detection \cite{BertiSpectroscopy2026}. In this section we derive the delay and check it against time-domain evolutions. The wave analysis below uses the effective model described in Table~\ref{tab:scope}: the lapse family and the closure introduced in Sec.~\ref{sec:echoes}.

\subsection{Setting}
\begin{figure}[!tbp]
\includegraphics[width=\columnwidth]{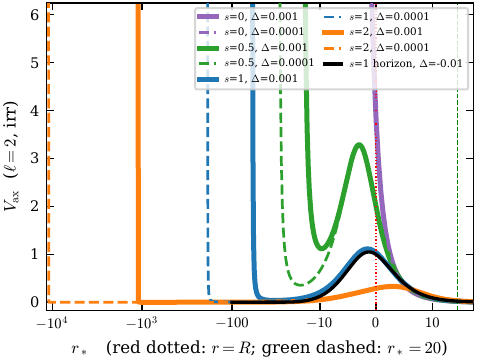}
\caption{The axial potential $V_{\rm ax}$ for $\ell=2$ against $r_*$, for $s=0,\tfrac12,1,2$ (colours) at $\Delta=10^{-3}$ (solid) and $10^{-4}$ (dashed), and the $s=1$ horizon run at $\Delta=-10^{-2}$ (black). The horizontal axis is $\mathrm{arcsinh}(r_*/5)$: linear near the barrier and logarithmic in the throat. The centrifugal wall sits at $r_*\simeq-\Lambda$ and moves out as $\Delta^{-s/2}$; the barrier depends on $s$ but not on $\Delta$; for $s=0$ there is no well. Red dotted: $r=R$; green dashed: $r_*=20$, beyond which the geometry is Schwarzschild.}
\label{fig:echo_pot}
\end{figure}

\begin{figure*}[!t]
\includegraphics[width=\textwidth]{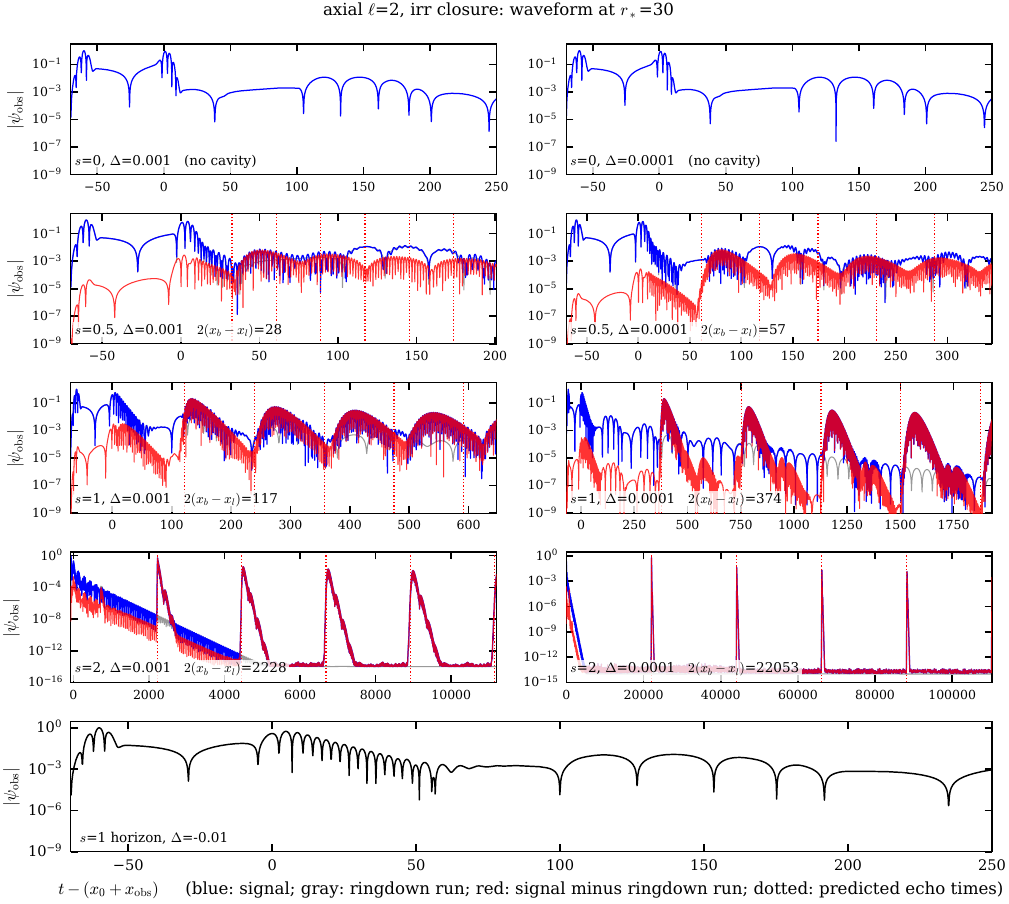}
\caption{Axial $\ell=2$ signal at $r_*=30$ against $t-(x_0+x_{\rm obs})$, for $s=0,\tfrac12,1,2$ (rows) and $\Delta=10^{-3},10^{-4}$ (columns), and the $s=1$ horizon run at $\Delta=-10^{-2}$ (bottom). Blue: signal; gray: ringdown run of the same $s$ ($\Delta=10^{-6}$, grid cut inside its cavity); red: their difference, i.e.\ the echoes alone. Dotted: reference marks $t_{\rm peak}+n\,2(x_b-x_l)$, measured from the prompt peak; the measured echoes lie later by about $n\delta$, shifted by the $\Delta$-independent first-gap offset of Sec.~\ref{sec:echotimes}, and Fig.~\ref{fig:echo_scal}b compares the spacings of successive echoes with the round trip. For $s=0$ no distinct echo is seen. The $s=2$ panels have an extended vertical range; values below about $10^{-13}$ of the prompt peak are round-off.}
\label{fig:echo_wave}
\end{figure*}

We evolve axial (odd-parity) perturbations. We use the irrotational closure, in which the axial covariant components of the fluid velocity and of the radial unit vector are unperturbed, $\delta u_\phi=\delta k_\phi=0$ (equivalently, the axial components of the perturbed effective stress vanish in the tetrad frame \cite{ChenChen2019}). Under this closure the perturbations obey Eq.~\eqref{eq:master} with
\begin{equation}
V_{\rm ax}=N^2\Bigl[\frac{\ell(\ell+1)}{r^2}-\frac{2(1-F)}{r^2}-\frac{F'/2+\Phi'F}{r}\Bigr],
\label{eq:Vax}
\end{equation}
which reduces to the Regge--Wheeler potential in vacuum \cite{ReggeWheeler1957,Chandrasekhar1983}. Using $1-F=2m/r$ and the field equations of Sec.~\ref{sec:lapse}, Eq.~\eqref{eq:Vax} can be written as $V_{\rm ax}=N^2[\ell(\ell+1)/r^2-6m/r^3+4\pi(\rho_E-p_r)]$. This is the axial potential of a perfect-fluid star \cite{ChandrasekharFerrari1991} with the pressure replaced by the radial stress, the form used for gravastars and other anisotropic sources \cite{ChirentiRezzolla2007,Boonserm2013}, and it is the irrotational branch of recent analyses of axial perturbations in anisotropic matter \cite{Chakraborty2025,DOnofrio2026,ZhaoPani2026}. The strictly static prescription of those analyses adds a term $-16\pi(p_r-p_t)N^2$ \cite{Chakraborty2025,ZhaoPani2026}. In that prescription the axial equation is generally coupled to the fluid perturbation, which we do not evolve; a dynamically consistent model that keeps this term must also evolve the axial displacement of the fluid \cite{RodriguezRuiz2026}. The two prescriptions agree at leading order when the anisotropy is small \cite{ZhaoPani2026}. Here, however, $|p_r-p_t|$ is of order $\rho_E$ (for $s=1$, $p_r-p_t=r\rho_E'/2$), so they need not agree. The ringdown frequencies and the echo amplitudes depend on the closure, whereas the leading delay, which involves only the optical length, does not. Since the matter has no specified dynamics, Eq.~\eqref{eq:Vax} is a modelling choice rather than a derivation, and we do not treat polar perturbations.

For $\Delta>0$ and $s>0$ the potential has three parts (Fig.~\ref{fig:echo_pot}), which correspond to the cap, the bulb and the neck of Fig.~\ref{fig:waveemb}. There is a centrifugal wall $\simeq\ell(\ell+1)N_0^2/r^2$ at the centre, a long shallow well in the throat where $V\propto N^2\sim\Delta^{s}$, and a barrier of height $V_b=O(1)$ at $r_*=x_b$, near $r=R$. The barrier depends on $s$ but not on $\Delta$; for $\ell=2$ its height is $V_b\simeq3.26$, $1.11$ and $0.33$ for $s=\tfrac12$, $1$ and $2$. For $s=0$ there is no well at all. The throat, where $V\simeq4N_0^2/R_M^2\approx34$, is then part of the barrier and forms no cavity for $\omega^2$ below this plateau.

In the numerical evolutions the waves are sent in from far away. We use $x\equiv r_*$, with $x=0$ at $r=R$, and follow the Burkert profile out to $x=20$, which corresponds to $r_c\simeq17.0R$, $16.1R$, $15.0R$ and $12.6R$ for $s=0,\tfrac12,1,2$. At that radius $2m/r\simeq0.12$--$0.15$ and still decreasing. Beyond $r_c$ we hold the mass fixed at $Mf(r_c)$, so that the far field is exactly Schwarzschild; this only removes the slow logarithmic growth of $m$ from the region where the waves propagate. Because $F'$ and the gradient of the lapse jump at $r_c$, the junction carries a thin layer with zero surface energy and a small tangential surface pressure, $|S|\lesssim1.5\times10^{-4}$ in units $R=1$, and $V_{\rm ax}$ jumps there by $0.4$--$1.3\%$. Its only visible effect on the waves is a weak reflection, and moving $r_c$ (see below) bounds its influence on the quoted results. Because the barrier lies inside the matter, the prompt ringdown below is that of this object and not that of Schwarzschild, as for black holes embedded in matter halos \cite{BarausseCardosoPani2014,CardosoGalaxies2022}. A Gaussian packet with $\omega_0=0.8$ and width $2.5$ starts at $x=60$ and travels inward, and the signal is recorded at $x=30$. We integrate $\psi_{tt}-\psi_{xx}+V_{\rm ax}\psi=0$ with a second-order leapfrog scheme ($\Delta x=0.04$, $\Delta t=0.02$), absorbing layers at the outer boundary and a reflecting centre for $\Delta>0$. We evolve $s=0,\tfrac12,1,2$ at $\Delta=10^{-3}$ and $10^{-4}$, and $s=\tfrac12,1$ also at $10^{-5}$ and $10^{-6}$. For $s=2$ at $\Delta\le10^{-5}$ the cavity is longer than $10^5$, too long to evolve, and we compute the echoes in the frequency domain instead (Appendix~\ref{app:fd}). The results are well converged. Halving the grid spacing ($\Delta x:0.04\to0.02$) changes the echo spacings by at most $0.4$, the echo amplitudes by at most $1\%$ and the fitted ringdown frequencies by at most $2\times10^{-3}$ for $s=\tfrac12$ and $2$ at $\Delta=10^{-3}$, and leaves the $s=1$ values at $\Delta=3\times10^{-3}$ unchanged to $10^{-4}$. Moving the outer radius from $x=20$ to $x=40$ changes nothing at the quoted precision for $s=1$, $\Delta=10^{-3}$, while moving it in to $x=10$ leaves the echo spacings unchanged, keeps the echo amplitudes within $0.3\%$ and shifts the ringdown frequencies by up to $2\%$. Signals below about $10^{-13}$ of the prompt peak are at the round-off floor. 

\subsection{Barrier plus cavity: echo times}
\label{sec:echotimes}
Consider a wave of frequency $\omega$ that hits the barrier from outside, and let $R_b(\omega)$ and $T_b(\omega)$ be its reflection and transmission amplitudes. For a wave hitting the barrier from the cavity side we write $\bar R_b$ and $\bar T_b$ (the bars do not denote derivatives). The centre is a perfect reflector with phase $\rho(\omega)$, $|\rho|=1$, and the cavity has optical half-length
\begin{equation}
L\equiv x_b-x_l=\int_{r_l}^{r_b}\frac{dr}{N\sqrt F},
\end{equation}
where $x_l$ is the inner end of the grid ($r_l\to0$). Adding up the multiple reflections between the barrier and the wall, as in the transfer-function description of echoes \cite{MarkZimmerman2017,MaggioTesta2019,CardosoPani2019}, gives the total reflection amplitude
\begin{equation}
R_{\rm tot}(\omega)=R_b+T_b\bar T_b\sum_{n\ge1}\bigl(\bar R_b\rho\,e^{2i\omega L}\bigr)^{n-1}\rho\,e^{2i\omega L}.
\label{eq:Rtot}
\end{equation}
The first term is the prompt signal; it involves only the barrier and therefore does not depend on $\Delta$. The $n$-th term of the sum is the $n$-th echo, with amplitude $|T_b\bar T_b||\bar R_b|^{n-1}$ and phase $n[2\omega L+\arg(\rho \bar R_b)]$. The spacing between successive echoes is the group delay of the round-trip phase, $t_{n+1}-t_n=T_{\rm echo}$, where
\begin{equation}
T_{\rm echo}(\omega)=\frac{d}{d\omega}\Bigl[2\omega L+\arg(\rho \bar R_b)\Bigr]=2\,\frac{d\Phi}{d\omega},
\label{eq:Techo-def}
\end{equation}
where $\Phi(\omega)=\int_{x_1}^{x_2}\sqrt{\omega^2-V}\,dx$ is the WKB phase between the turning points $x_1$ (centrifugal wall) and $x_2$ (barrier), which exist for $\omega^2<V_b$. At fixed frequency, the first echo trails the prompt signal by the same $T_{\rm echo}$. For a real potential the Wronskians of the two scattering solutions give $\arg(R_b\bar R_b)=\arg(T_b\bar T_b)+\pi$ at every $\omega$, whatever reference points are used for the amplitudes, so the transmissions carry no group delay beyond that of the reflections. Any first-gap offset must therefore come from the frequency dependence of the phases of the barrier and the wall. The prompt signal and the echoes sample these phases at different dominant frequencies, because $|R_b|$ and $|T_b\bar T_b|$ weight the packet spectrum differently: the prompt signal is dominated by frequencies below the cutoff frequency $f$ of Sec.~\ref{sec:greybody}, the echoes by frequencies above it. In the evolutions the first gap is longer than the later spacings for $s=\tfrac12$ and shorter for $s=1$ and $2$, by $10\%$, $8\%$ and $1.5\%$ of the spacing at $\Delta=10^{-3}$ and by $2\%$, $0.3\%$ and $0.15\%$ at the smallest evolved $\Delta$; the first-gap offset itself does not depend on $\Delta$, so its fraction of the spacing falls as $\Delta^{s/2}$. Equivalently, the poles of Eq.~\eqref{eq:Rtot}, where $\bar R_b\rho\,e^{2i\omega L}=1$, are quasi-bound cavity frequencies spaced by $\delta\omega=2\pi/T_{\rm echo}$. For $\omega^2<V_b$, differentiating $\Phi$ gives
\begin{equation}
T_{\rm echo}(\omega)=2\int_{x_1}^{x_2}\frac{\omega\,dx}{\sqrt{\omega^2-V(x)}} .
\label{eq:Techo-wkb}
\end{equation}

\begin{proposition}[echo delay]\label{prop:echo}
Let $s>0$ and let $\omega>0$ be fixed. As $\Delta\to0$, the spacing of successive echoes is
\begin{equation}
T_{\rm echo}(\omega)=2\Lambda(\Delta,s)+\delta(\omega)+o(1),
\label{eq:Techo}
\end{equation}
where $\Lambda$ is the optical length of the throat and $\delta(\omega)=O(1)$ depends on the barrier region and the centrifugal wall but not on $\Delta$. For $0<s<2$, $\Lambda$ differs from the leading term of Eq.~\eqref{eq:Lambda} by a $\Delta$-independent constant, which can be absorbed into $\delta$, so that
\begin{equation}
T_{\rm echo}\simeq
\begin{cases}
\dfrac{2\pi}{N_0\sqrt{\beta\Delta}}+\delta, & s=1,\\[3mm]
\dfrac{2c_s\,\Delta^{-s/2}}{N_0\sqrt\beta}+\delta, & 0<s<2,
\end{cases}
\end{equation}
and the echo spacing diverges as $\Delta^{-s/2}$. For $s\ge2$ the next terms in the expansion of $F$ about its minimum add to $2\Lambda$ a term $2C\ln(1/\Delta)$ for $s=2$ ($C\simeq2.18$ for the Burkert profile) and powers $\Delta^{1-s/2}$ for $s>2$, which must then be kept. For $\omega^2<V_b$, $\delta$ follows from Eq.~\eqref{eq:Techo-wkb}. For $\omega^2>V_b$ the barrier has no turning point, $\bar R_b$ is an above-barrier reflection amplitude, and $\delta$ includes its group delay. For $s=0$ the throat forms no cavity below its plateau, and Eq.~\eqref{eq:Techo} does not apply.
\end{proposition}
\emph{Proof sketch.} Write the round-trip phase as $\Theta=2\int k\,dx+\arg\rho+\arg\bar R_b$, where $k=\sqrt{\omega^2-V}$ and the integral runs over the region between the wall and the barrier in which $V<\omega^2$; then $T_{\rm echo}=d\Theta/d\omega$. Most of this region is the throat, where $V=N^2W$ with $W=[\ell(\ell+1)-2(1-F)]/r^2+\cdots$ smooth and $N^2\propto F^s$ small. There $V$ varies over distances of order $\Lambda$ in $x$, so WKB holds for every fixed $\omega$, including $\omega^2>V_b$. We split the throat's contribution into the free travel time and a correction, $2\int\omega\,dx/k=2\Lambda_{\rm th}+2\int[\omega/k-1]\,dx$, where $\Lambda_{\rm th}$ is the optical length of the throat. The correction is $\simeq\int V\,dx/\omega^2$, and since $V\,dx=N\,W\,dr/\sqrt F$ with $N/\sqrt F\propto F^{(s-1)/2}\sim|\zeta|^{s-1}$, which is integrable at $\zeta=0$ for $s>0$, it converges to a $\Delta$-independent limit as $\Delta\to0$ (Appendix~\ref{app:echo}). The remaining contributions, namely the wall phase $\rho$, the reflection phase of $\bar R_b$ (by tunnelling for $\omega^2<V_b$, above the barrier for $\omega^2>V_b$) and $2(\Lambda_{\rm th}-\Lambda)$, come from regions where $F$ and $N$ are $O(1)$, so they do not depend on $\Delta$ either; together they define $\delta(\omega)$. For $\omega^2<V_b$ the same result follows directly from Eq.~\eqref{eq:Techo-wkb}. $\square$

\paragraph*{Comparison with horizon-scale echoes.}
Consider a Schwarzschild exterior with a reflecting surface at proper distance $d$ outside the would-be horizon. There the delay grows only logarithmically, $T_{\rm echo}\sim4M|\ln(d^2/M^2)|$ \cite{CardosoFranzinPani2016,CardosoHopper2016,AbediDykaarAfshordi2017,CardosoPani2019} (Eq.~(3.48) of Ref.~\cite{BertiSpectroscopy2026}), because the redshift diverges at a simple zero of $F$; for a Planckian $d$ this delay is of the order of the scrambling time \cite{AbediDykaarAfshordi2017,SaraswatAfshordi2020}. Thin-shell gravastars and wormholes close to a would-be horizon behave in the same way \cite{CardosoHopper2016,DamourSolodukhin2007}. In our case the zero of $F$ is double, and the delay is set by the lapse instead. For $s>0$ it is a power law, $\Delta^{-s/2}$, the kind of long, geometric delay that has been discussed for exotic interiors \cite{ZimmermanGeorgeChen2023,ArrecheaLiberatiVellucci2024} (see Sec.~3.3.4 of Ref.~\cite{BertiSpectroscopy2026}); for $s=0$ the throat is part of the barrier and forms no cavity at these frequencies. Writing $N_{\rm min}=N_0\Delta^{s/2}$ for the lapse at the bottom of the throat, the delay is simply the inverse redshift there, $T_{\rm echo}\propto1/N_{\rm min}$, while the proper length of the throat grows only as $\ln(1/\Delta)$. The capped throats of microstate geometries show the same separation of scales: the echo period and the trapping times scale as the inverse redshift of the cap, while the length of the throat grows only logarithmically \cite{Bena2019Thermal,Bena2021Escape}.

For $s=1$ the delay can also be read off from the near-throat spectrum. To leading order the throat is global AdS$_2$, with $\omega_{n\ell}=N_0\sqrt{\beta\Delta}\,(n+h_\ell)$ (Sec.~\ref{sec:lapse}; for axial waves $h_\ell$ contains $(\ell-1)(\ell+2)$). The level spacing is $\delta\omega=N_0\sqrt{\beta\Delta}=\pi/\Lambda$, so $2\pi/\delta\omega=2\Lambda$, in agreement with Eq.~\eqref{eq:Techo}. Correspondingly, the spectrum of the reflected signal shows an echo comb with teeth spaced by $\delta\omega=2\pi/T_{\rm echo}$.

\paragraph*{Amplitudes.} According to Eq.~\eqref{eq:Rtot}, the $n$-th echo has amplitude $|T_b\bar T_b||\bar R_b|^{n-1}$ times the spectral weight of the packet, and this depends only on the barrier. The echo amplitudes should therefore approach $\Delta$-independent values as $\Delta\to0$. Successive echoes do not shrink by a fixed factor $|\bar R_b|$, however, because $\bar R_b$ depends on frequency and each pass filters the packet. The barrier transmits high frequencies and reflects low ones (Sec.~\ref{sec:greybody}). Each successive echo has been reflected inside the barrier one more time, so the echoes are dominated by progressively lower frequencies \cite{CardosoFranzinPani2016,MaggioTesta2019,BertiSpectroscopy2026}. In horizon-scale models the role of $|\bar R_b|$ and of the perfectly reflecting centre is played by the reflectivity of the inner boundary \cite{CardosoPani2019,WangAfshordi2018,OshitaWangAfshordi2020,WangOshitaAfshordi2020,MaggioCardosoDolanPani2019}. How large the echoes are is set by the height of the barrier. For $s=\tfrac12$ the barrier ($V_b\simeq3.3\gg\omega_0^2$) is a strong mirror, $|T_b\bar T_b|$ is small, and the echoes are only about $1\%$ of the prompt signal. For $s=1$ ($V_b\simeq\omega^2$ at the dominant ringdown frequency) the barrier is a weak mirror, and the first echo reaches $0.38$ of the prompt peak. For $s=2$ ($V_b\simeq0.33<\omega_0^2$) the packet lies above the barrier, which barely reflects it. The prompt signal is then weak and the first echo is about twice as large, while the next echo is an order of magnitude smaller ($\simeq0.08$), set by the small above-barrier reflection $|\bar R_b|$. Later ratios ($0.5$--$0.65$) are larger, because each pass shifts the echo to lower frequencies, where $|\bar R_b|$ is larger. Unlike the leading delay, the amplitudes depend on the closure through the barrier.

\paragraph*{Power-law decay of the echo train.} From the second echo on, the amplitudes do not fall geometrically but as a power of the echo number (Fig.~\ref{fig:echo_scal}c),
\begin{equation}
A_n\propto n^{-p},
\label{eq:echo_powerlaw}
\end{equation}
with $p\simeq1.05$, $1.30$ and $1.40$ for $s=\tfrac12$, $1$ and $2$. These values come from fits over echoes $2$--$40$, which we follow in the frequency domain (Appendix~\ref{app:fd}) at $\Delta=10^{-3}$. The fits leave residuals of $0.01$--$0.02$ in $\ln A_n$, whereas an exponential in $n$ leaves residuals of $0.27$--$0.38$, which are $15$--$40$ times larger. The evolved echoes $2$--$4$ give $p=0.96$--$1.03$, $1.38$ and $1.58$, and the local index $-d\ln A_n/d\ln n$ drifts slowly towards the values above. For $s=1$ and $2$ the first echo lies above the line. It has crossed the barrier twice but has never been reflected inside it, so it still carries the frequencies above the barrier.

The power law can be understood from Eq.~\eqref{eq:Rtot} and the greybody factor of Sec.~\ref{sec:greybody}. With flux-normalized amplitudes, the spectral weight of the $n$-th echo is $|T_b\bar T_b||\bar R_b|^{n-1}=\Gamma(1-\Gamma)^{(n-1)/2}$. As a function of $\Gamma$, this is largest at $\Gamma=2/(n+1)$, where for large $n$ it is $\simeq2/[e(n+1)]$. Intuitively, frequencies with $\Gamma\gg1/n$ have already leaked out of the cavity, and those with $\Gamma\ll1/n$ have hardly got in. Below the barrier $\Gamma$ grows exponentially with $\omega$, so the frequencies with $\Gamma\simeq1/n$ form a band of fixed width that moves down by a fixed step each time $n$ doubles. For $s=1$, for example, the amplitude-weighted mean frequency of echoes $1$, $2$, $5$, $10$ and $40$ is $1.28$, $1.09$, $1.01$, $0.97$ and $0.92$. The spectral weight in this band therefore falls as $1/n$, and indeed summing the spectra of the echoes without their phases gives smaller exponents than measured, $p\approx0.7$--$0.8$, $1.0$--$1.1$ and $1.05$--$1.2$ for $s=\tfrac12$, $1$ and $2$, depending on how the spectral weight is measured. The rest of $p$ comes from the phases. The group delay of a round trip varies across the band, so the echoes spread out in time as $n$ grows, and the peak of the envelope falls faster than the spectral weight. The exponent therefore depends on both the barrier and the spectrum of the incident packet; the values quoted above are for the packet with $\omega_0=0.8$. An echo train whose successive ratios approach $1$ in this way is a signature of a perfectly reflecting interior behind a frequency-dependent barrier. A partially absorbing interior, with reflectivity $|\rho|<1$ at all frequencies, would add a geometric factor $|\rho|^n$ and eventually turn the power law into an exponential decay.

\subsection{Numerical results}\label{sec:numres}
\begin{figure*}[!t]
\includegraphics[width=\textwidth]{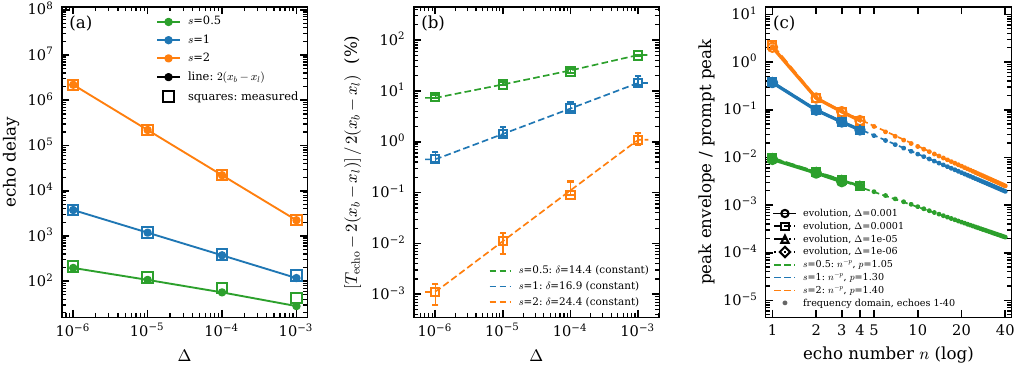}
\caption{(a) Echo spacing against $\Delta$ for $s=\tfrac12,1,2$: the round trip $2(x_b-x_l)$ computed from the background (lines, not an evolution; fitted slopes $-0.28$, $-0.50$ and $-1.00$, against $-s/2$) and the spacing of successive echoes measured in the evolutions (squares) and, for $s=2$ at $\Delta\le10^{-5}$, in the frequency-domain computation of Appendix~\ref{app:fd}. (b) Excess of the measured spacing over the round trip, $[T_{\rm echo}-2(x_b-x_l)]/2(x_b-x_l)$ in percent; squares: median over the spacings of echoes $1$--$4$, bars: their range; dashed: a constant offset $\delta$ (the median over all $\Delta$), which gives lines $\propto\Delta^{s/2}$. (c) Smoothed envelope of the echoes relative to the prompt peak against the echo number, on logarithmic axes: echoes $1$--$4$ for all evolved $(s,\Delta)$ (curves with the same $s$ coincide, i.e.\ the echo amplitudes are independent of $\Delta$), echoes $1$--$40$ from the frequency-domain computation at $\Delta=10^{-3}$ (dots), and the power laws $n^{-p}$ of Eq.~\eqref{eq:echo_powerlaw} fitted to echoes $2$--$40$ (dashed).}
\label{fig:echo_scal}
\end{figure*}

\begin{figure*}[!t]
\includegraphics[width=\textwidth]{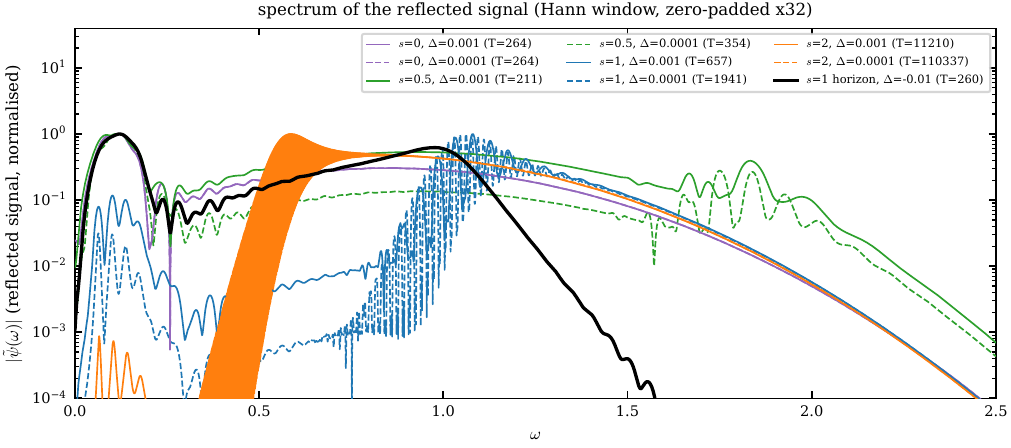}
\caption{Spectrum of the signal reflected from the object (Hann window, zero-padded), each curve normalised to its maximum, for $s=0,\tfrac12,1,2$ at $\Delta=10^{-3}$ (solid) and $10^{-4}$ (dashed), and the $s=1$ horizon run (black, thick; drawn on top). Above $\omega\simeq1$ the horizon spectrum falls steeply because the transmitted wave is absorbed, while every $\Delta>0$ spectrum keeps the shape of the incident Gaussian because the wave is reflected at the centre. The $s=2$ comb (spacing $2\pi/T_{\rm echo}\simeq3\times10^{-3}$) is not resolved at this scale.}
\label{fig:echo_spec}
\end{figure*}

For $\ell=2$ and the Burkert profile, the geometric part of Proposition~\ref{prop:echo} can be checked without any evolution. For $s=1$, $1.5$ and $2$ the cavity half-length $x_b-x_l$, measured to the barrier maximum, agrees with $\Lambda$ of Eq.~\eqref{eq:Lambda} to $1.0\%$, $1.6\%$ and $1.2\%$ at $\Delta=10^{-3}$, and to $0.1\%$ at $\Delta=10^{-5}$ for $s=1$ and $1.5$. For $s=0.5$ the ratio $(x_b-x_l)/\Lambda$ is $0.78$ at $\Delta=10^{-3}$ and $0.93$ at $10^{-5}$, consistent with corrections of order $\Delta^{s/2}$. For $s=0$ the throat has no barrier maximum.

To isolate the echoes, we subtract from each signal a ringdown run with the same $s$ at $\Delta=10^{-6}$. The grid of this reference run is cut inside its own cavity, at $60\%$ of the cavity length or at $r_*=-400$, and closed with an absorbing layer, so its centrifugal wall is removed and no echo can come back. Near the barrier its potential coincides with that of the $\Delta>0$ runs up to $O(\Delta)$, so the difference contains the echoes but no prompt signal, for every $s$. (A horizon run at $-\Delta$ would serve the same purpose only for $s=1$, since $\Delta<0$ is singular for $s\ne1$.) For $s=1$, cutting the grid changes the ringdown run by at most $1.5\times10^{-5}$ of the prompt peak, compared with the full-cavity run before its own first echo.

\paragraph*{Spacings.} Figure~\ref{fig:echo_scal}b shows, in percent, how much the measured spacing of successive echoes (echoes $1$--$4$) exceeds the cavity round trip $2(x_b-x_l)$. If the round-trip offset $\delta=T_{\rm echo}-2(x_b-x_l)$ does not depend on $\Delta$, as Proposition~\ref{prop:echo} states, the points should lie on lines $\propto\Delta^{s/2}$, and this is what we find. In the evolutions the round-trip offset is $\delta=12.9$--$16.5$, $15.1$--$23.4$ and $19.2$--$36.5$ for $s=\tfrac12$, $1$ and $2$, and the frequency-domain computation gives $13.8$--$34.9$ for $s=2$ at $\Delta=10^{-5}$ and $10^{-6}$. None of these shows a trend in $\Delta$ over three decades. The spread within each case reflects the frequency dependence of $\delta(\omega)$, since each pass through the barrier shifts the dominant frequency of the echo. As a fraction of the spacing, the round-trip offset falls from $49$--$51\%$, $13$--$20\%$ and $0.9$--$1.5\%$ at $\Delta=10^{-3}$ to $7.2$--$8.4\%$, $0.41$--$0.63\%$ and $0.0006$--$0.0016\%$ at $\Delta=10^{-6}$. The leading-order estimate $2\Lambda$ of Eq.~\eqref{eq:Lambda} agrees with the numerical round trip to $4\%$, $0.02\%$ and $0.005\%$ at $\Delta=10^{-6}$. For $s=1$, Eq.~\eqref{eq:Techo-wkb} at $\omega=1$ exceeds $2\Lambda$ by $10.7\%$, $3.5\%$, $1.1\%$ and $0.35\%$ at $\Delta=10^{-3}$, $10^{-4}$, $10^{-5}$ and $10^{-6}$, against $11.7$--$18.4\%$, $3.8$--$5.8\%$, $1.2$--$1.9\%$ and $0.38$--$0.60\%$ measured, so it reproduces the measured spacings to within $6.5\%$, $2.2\%$, $0.7\%$ and $0.25\%$.

We fit the prompt ringdown of the ringdown runs with a matrix-pencil method \cite{BertiMatrixPencil2007}. The fitted modes agree between $\Delta=10^{-3}$ and $10^{-4}$ to $3\times10^{-3}$, confirming that the prompt signal is set by the barrier, as Eq.~\eqref{eq:Rtot} says. They do depend on $s$, because the lapse multiplies the potential, as one expects when matter sits near the barrier \cite{BarausseCardosoPani2014,CheungElephantFlea2022,BertiStability2022,CardosoGalaxies2022}.

After the ringdown, echoes arrive at intervals $T_{\rm echo}$ for $s=\tfrac12$, $1$ and $2$ (Fig.~\ref{fig:echo_wave}). The horizon run shows none, and for $s=0$ no distinct echo appears for this packet, whose $\omega_0^2=0.64$ lies far below the throat plateau $V\simeq34$. The round trip $2(x_b-x_l)$ scales as $\Delta^{-0.28}$, $\Delta^{-0.50}$ and $\Delta^{-1.00}$ for $s=\tfrac12$, $1$ and $2$ (Fig.~\ref{fig:echo_scal}), compared with the expected $\Delta^{-s/2}$. For $s=\tfrac12$ the difference comes from the $O(\Delta^{s/2})$ correction to $\Lambda$. Indeed, fitting the round trip to $a\Delta^{-s/2}+b$ over $10^{-6}\le\Delta\le10^{-3}$ gives $a=6.475$, $3.7509$ and $2.2017$ for $s=\tfrac12$, $1$ and $2$, against $2c_s/(N_0\sqrt\beta)=6.473$, $3.7505$ and $2.2021$, and for $s=\tfrac12$ the constant $b\simeq-8.1$ explains the slope $-0.28\pm0.006$ of a pure power-law fit. Relative to the prompt peak, the smoothed envelopes of echoes $1$--$4$ are $0.0096$, $0.0049$, $0.0034$ and $0.0025$ for $s=\tfrac12$; $0.38$, $0.099$, $0.056$ and $0.038$ for $s=1$; and $2.1$--$2.3$, $0.18$, $0.091$ and $0.059$ for $s=2$. They agree to within $5\%$ over $10^{-6}\le\Delta\le10^{-3}$, and to within $3\%$ for $\Delta\le10^{-4}$. The one exception is the first $s=2$ echo ($2.05$ against $2.26$), whose reference, the weak prompt peak, is itself sensitive. This is the $\Delta$-independence expected from Eq.~\eqref{eq:Rtot}.

\paragraph*{Spectrum of the reflected signal.}
The Fourier transform of the signal at $r_*=30$ (Fig.~\ref{fig:echo_spec}) distinguishes reflecting from absorbing interiors at high frequency. The reflected spectrum is the incident spectrum multiplied by a reflection coefficient, $|\tilde\psi_{\rm refl}(\omega)|=|R(\omega)|\,|\tilde\psi_{\rm inc}(\omega)|$, and the incident packet is a Gaussian in $\omega$ centred at $\omega_0=0.8$ with width $0.4$. Above the barrier frequency most of the wave is transmitted. In the horizon run the transmitted part falls into the horizon and is lost, so $|R(\omega)|$ drops steeply: the spectrum falls by a factor of about $3\times10^{-4}$ between $\omega\simeq1$ and $1.55$, and only the low-frequency part reflected by the barrier survives. In every $\Delta>0$ run, $s=0$ included, nothing is absorbed. Whatever crosses the barrier is reflected at the throat plateau or at the centre and comes back out, so $|R_{\rm tot}(\omega)|=1$ in Eq.~\eqref{eq:Rtot}. The spectrum of the full signal then keeps the shape of the incident Gaussian, and its high-frequency slope is set by the width of the packet, not by the object. A slow high-frequency decline is therefore the signature of a reflecting interior, and the fast decline of the horizon spectrum that of an absorbing one (Sec.~\ref{sec:greybody} quantifies both). Frequency-dependent reflectivities, such as those proposed for quantum black-hole horizons \cite{OshitaWangAfshordi2020,WangOshitaAfshordi2020}, would interpolate between the two cases. For $s>0$ the echoes also add a comb with spacing $2\pi/T_{\rm echo}$ \cite{MarkZimmerman2017,WangAfshordi2018}, which is visible for $s=1$ near $\omega\simeq1$ (spacing $0.047$ at $\Delta=10^{-3}$ and $0.016$ at $10^{-4}$). Each curve is normalised to its own maximum, so vertical offsets between the curves have no physical meaning.

\subsection{Greybody factor of the barrier}
\label{sec:greybody}
\begin{figure*}[!t]
\includegraphics[width=\textwidth]{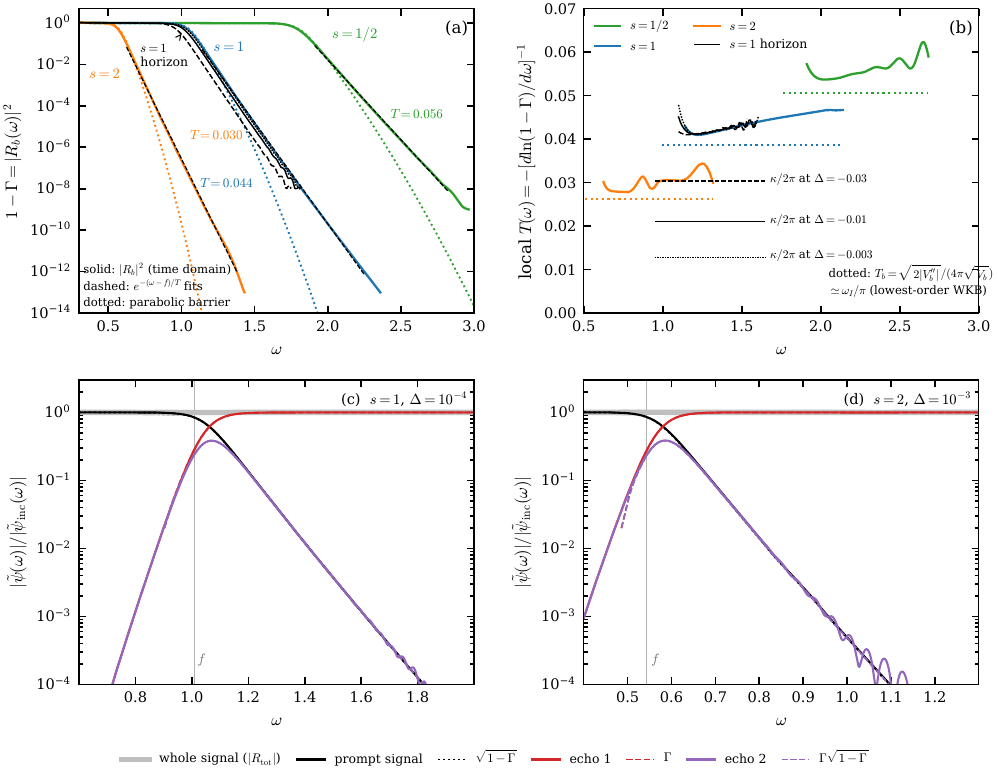}
\caption{Greybody factor of the barrier ($\ell=2$, units $R=1$). (a) Reflectivity $1-\Gamma=|R_b|^2$ of the barrier alone: ringdown runs ($\Delta=10^{-6}$, throat cut by an absorbing layer) for $s=\tfrac12,1,2$ (colours), and $s=1$ horizon runs at $\Delta=-0.003$ (dotted), $-0.01$ (solid) and $-0.03$ (dashed, black). Black dashed lines: fits of Eq.~\eqref{eq:greybody}, with the slopes $T$ marked; coloured dotted curves: parabolic barrier, Eq.~\eqref{eq:fermi}. (b) Local slope $T(\omega)=-[d\ln(1-\Gamma)/d\omega]^{-1}$; coloured dotted lines: $T_b$ of Eq.~\eqref{eq:fermi}; black horizontal lines: $\kappa/2\pi$ of the horizon runs. (c), (d) Spectra of the prompt signal, the first two echoes and the whole signal of runs with a cavity, divided by the incident spectrum, with the predictions $\sqrt{1-\Gamma}$, $\Gamma$, $\Gamma\sqrt{1-\Gamma}$ and $1$ of Eq.~\eqref{eq:Rtot}, $\Gamma$ taken from panel (a) and shown where $\Gamma>0.02$. The whole-signal curve is averaged over the echo comb; the grey vertical line marks $f$.}
\label{fig:greybody}
\end{figure*}

The barrier divides the incident spectrum between the prompt signal and the echoes, and how it does so is set by its greybody factor, the transmission probability $\Gamma(\omega)=1-|R_b(\omega)|^2$. Since the incident packet is known, $|R_b|$ can be measured directly as the ratio of the reflected and incident spectra at $r_*=30$. We do this for the ringdown runs, in which an absorbing layer inside the throat removes whatever crosses the barrier, and for $s=1$ horizon runs at $\Delta=-0.003$, $-0.01$ and $-0.03$. The record is filtered to remove the slow far-field signal of the packet's low-frequency content, and the reflected part is taken after the weak reflection from $r_c$, where the slopes of $F$ and of the lapse change because the mass is held fixed beyond it. To follow the high-frequency tail we add packets with $\omega_0=1.6$ and $2.4$. Where they overlap, the three packets give the same $|R_b|$, and halving the grid spacing changes the slopes below by less than $10^{-4}$.

Above the barrier frequency the reflectivity falls exponentially (Fig.~\ref{fig:greybody}a),
\begin{equation}
1-\Gamma(\omega)\simeq e^{-(\omega-f)/T},
\label{eq:greybody}
\end{equation}
which is the form Oshita found for the ringdown of Kerr black holes \cite{Oshita2024}. A linear fit of $\ln(1-\Gamma)$ above $1.1f$ gives $T=0.056$, $0.044$ and $0.030$ for $s=\tfrac12$, $1$ and $2$, over $7$, $11$ and $11$ decades of $1-\Gamma$. The cutoff frequencies are $f=1.77$, $1.01$ and $0.54$, close to the real parts $1.74$, $1.04$ and $0.57$ of the least-damped prompt modes. The three horizon runs give $T=0.0426$, $0.0424$ and $0.0421$ over $6$ decades, below which the absorbing layer sets a floor; since the fitted $T$ grows slowly with the width of the window, these values should be compared with the ringdown runs at the same frequencies, as in Fig.~\ref{fig:greybody}b, rather than with the 11-decade fits.

The exponential can be traced to the shape of the barrier near its top. For an inverted parabola, $V\simeq V_b-\tfrac12|V_b''|(r_*-x_b)^2$, the reflectivity is a Fermi function \cite{IyerWill1987},
\begin{equation}
\begin{gathered}
1-\Gamma=\Bigl[1+e^{2\pi(\omega^2-V_b)/\sqrt{2|V_b''|}}\Bigr]^{-1}\simeq e^{-(\omega-\sqrt{V_b})/T_b},\\
T_b=\frac{\sqrt{2|V_b''|}}{4\pi\sqrt{V_b}},
\end{gathered}
\label{eq:fermi}
\end{equation}
and at the same WKB order the fundamental mode has $\omega_I=\pi T_b$ \cite{SchutzWill1985}. The width of the cutoff and the damping of the prompt ringdown thus come from the same curvature of the barrier; this is the correspondence between greybody factors and quasinormal modes discussed in Refs.~\cite{KonoplyaZhidenko2024,Oshita2024}. Here $T_b=0.051$, $0.039$ and $0.026$. Well above the top, the parabola underestimates the reflectivity (dotted curves in Fig.~\ref{fig:greybody}a). The true decay stays exponential in $\omega$ rather than in $\omega^2$, and the local slope $T(\omega)=-[d\ln(1-\Gamma)/d\omega]^{-1}$ lies $10$--$15\%$ above $T_b$ and rises slowly with $\omega$ (Fig.~\ref{fig:greybody}b). Schwarzschild behaves the same way. For $\ell=2$ the exact Regge--Wheeler reflectivity gives $T=0.067/2M$ on the frequency window of Ref.~\cite{Oshita2024}, reproducing the value found there, and this is $1.18$ times $\omega_I/\pi$.

The results indicate that the high-frequency cutoff is determined by the barrier and is insensitive to the interior in the configurations studied here. The three horizon runs have $\kappa/2\pi=0.013$, $0.021$ and $0.030$, yet their slopes agree with each other to $1\%$, and they agree at the same frequencies with the horizonless ringdown run of the same $s$ (Fig.~\ref{fig:greybody}b). The prompt signal and its spectrum therefore cannot tell a horizon from a reflecting throat, whose surfaces in Fig.~\ref{fig:waveemb} differ only below the neck: both carry the factor $\sqrt{1-\Gamma}$.

Where they differ is in what happens to the transmitted part: it either disappears into the open funnel of Fig.~\ref{fig:waveemb}b or enters the closed bulb of Fig.~\ref{fig:waveemb}a and comes back. In Eq.~\eqref{eq:Rtot}, with flux-normalized amplitudes, the prompt signal is $|R_b|=\sqrt{1-\Gamma}$, the first echo $|T_b\bar T_b|=\Gamma$ and the $n$-th echo $\Gamma(1-\Gamma)^{(n-1)/2}$, and for a perfectly reflecting centre the whole signal has $|R_{\rm tot}|=1$. Figures~\ref{fig:greybody}c and~d test these relations on runs with $s=1$, $\Delta=10^{-4}$ and $s=2$, $\Delta=10^{-3}$, in which the echoes are well separated in time. After division by the incident spectrum, the spectra of the prompt signal and of the first two echoes follow $\sqrt{1-\Gamma}$, $\Gamma$ and $\Gamma\sqrt{1-\Gamma}$, with $\Gamma$ taken from the ringdown runs; above $f$ the median deviations are $0.1$--$2\%$. Averaged over the echo comb, the whole signal has $|R_{\rm tot}|^2=1.002$ and $1.004$. Below $f$ the prompt signal carries most of the spectrum, while above $f$ the barrier transmits nearly everything and the first echo carries it instead. For a horizon, the high-frequency spectrum is suppressed according to Eq.~\eqref{eq:greybody}, a cutoff that has been proposed as a ringdown observable that is stable under small changes of the potential \cite{RosatoDestounisPani2024,Oshita2024}. For the throat, the echoes restore it.

Nearly extremal black holes have zero-damping modes, near-horizon modes whose decay rate vanishes at extremality \cite{YangZimmerman2013,ZimmermanMark2016}; the AdS$_2$ throat of Sec.~\ref{sec:lapse} plays the role of that near-horizon region, but it does not set the cutoff here, since $T$ does not depend on $\kappa$. In nearly extremal Kerr, rotation brings the near-horizon modes to $m\Omega_H$, close to the barrier frequency of the $\ell=m$ mode, and the values of $T_{22}$ in Ref.~\cite{Oshita2024} approach the Hawking temperature as the spin approaches $1$.

\subsection{Limits of this analysis}
\label{sec:stability}

The results are subject to the following restrictions. The ringdown and the echo amplitudes are properties of this particular matter distribution, not of Schwarzschild. Polar perturbations require a matter model and are not computed. The evolutions cover $\ell=2$ and $s=0,\tfrac12,1,2$; for other values of $s$ we check the dependence only through the geometry ($x_b-x_l\simeq\Lambda$), and for $s=2$ at $\Delta\le10^{-5}$ the echoes are computed in the frequency domain rather than evolved. For $\Delta<0$ only $s=1$ gives a regular horizon, so the horizon comparison is made for $s=1$ alone. For $s\neq1$ the effective source violates the null energy condition on one side of the throat (Sec.~\ref{sec:lapse}). The constant $\delta$ is measured, and for $s=1$ also evaluated from Eq.~\eqref{eq:Techo-wkb} at $\omega=1$ (Sec.~\ref{sec:numres}), but we have not derived a closed form for it. Finally, a perfectly reflecting centre traps waves for a long time. For horizonless objects with stably trapped modes or a stable inner light ring, this has been argued to lead to nonlinear instabilities \cite{Keir2016,EperonReallSantos2016,Cunha2017,GuoQBH2022}, and for microstate geometries tidal effects limit how deep into the throat a classical description can be trusted \cite{Tyukov2018}. Our analysis is restricted to linear perturbations.

\section{Conclusions}
\label{sec:conclusions}

\emph{Geometry (Level I).} For a prescribed smooth family in which a nondegenerate minimum of $F$ reaches zero, the proper radial length, the spatial volume and, on the connected subcritical slice, the geodesic length across the throat all show the same logarithmic critical scaling,
\begin{equation}
\mathcal{L},\mathcal{V}\propto-\ln|\Delta|,\qquad\mathcal{D}\propto-\ln\Delta,
\end{equation}
with a coefficient fixed by the curvature of the minimum, $1/\sqrt\beta$ for $\mathcal L$, as the numerics confirm for the Burkert profile. The critical slice develops an asymptotically cylindrical end whose local curvature, $2/R_M^2$, stays finite. We call $M_c$ a geometric marginal-trapping threshold, because the limiting minimal sphere lies at infinite proper distance, and we make no claim about an apparent horizon without a specified extension or matter dynamics.

\emph{Lapse (Level II).} The slice does not fix the lapse, which enters the throat only through the optical length $\Lambda$, the tangential pressure and $\Nmin$. For $N\propto F^{s/2}$ we find $\Lambda\sim\ln(1/\Delta)$ for $s=0$ and $\Lambda\sim\Delta^{-s/2}$ for $s>0$, and near the threshold the throat is AdS$_2$ of radius $1/(s\sqrt\beta)$. For $s=1$ the near-throat region is global AdS$_2$ to leading order, and its asymptotic spectrum $\omega_{n\ell}=N_0\sqrt{\beta\Delta}(n+h_\ell)$ matches the numerical spectrum for the Burkert profile for $\ell\ge1$. At $\Delta=0$ the critical surface becomes a degenerate Killing horizon for $s>0$ and a horizonless cylinder for $s=0$. For all $s$ the curvature is bounded in the static frame and the Kretschmann scalar stays finite, but at the critical horizon the curvature seen by an infalling observer is finite only for $s=1$ or $0<s\le\tfrac12$ (Proposition~\ref{prop:curv}). For $s=1$ ($N^2=F$) the stresses take the familiar $g_{tt}g_{rr}=-1$ form, $p_r=-\rho_E$ and $p_t=-\rho_E-r\rho_E'/2$, and the null, weak and dominant energy conditions all hold, whereas every $s\neq1$ violates the null energy condition on one side of the throat.

\emph{Echoes (effective model).} For axial gravitational perturbations under the irrotational closure, the throat behind the barrier acts as a cavity when $s>0$ (Sec.~\ref{sec:echoes}). The prompt ringdown is set by the barrier: it depends on $s$ but not on $\Delta$. Successive echoes are spaced by $T_{\rm echo}\simeq2\Lambda+\delta\propto\Delta^{-s/2}$, where neither $\delta$ nor the echo amplitudes depend on $\Delta$. The amplitudes fall as a power of the echo number, $n^{-p}$ with $p\simeq1$--$1.4$, because each echo is dominated by the frequencies whose transmission probability is about $1/n$. The spacing law holds for packets above as well as below the barrier. For $s=0$ the throat is part of the barrier, and the packets we studied show no distinct echo. We derived these results and checked them by time-domain evolution for $\ell=2$, with $s=\tfrac12,1$ over $10^{-6}\le\Delta\le10^{-3}$ and $s=0,2$ over $10^{-4}\le\Delta\le10^{-3}$, and for $s=2$ down to $\Delta=10^{-6}$ with a frequency-domain computation. These calculations show that the model can produce echoes from a horizonless, matter-supported configuration with a capped near-AdS$_2$ throat (namely a quasi-black hole reviewed in Refs.~\cite{CardosoPani2019,AbediReview2020,BertiSpectroscopy2026}). Its predicted echo delay scales as $\Delta^{-s/2}$, rather than logarithmically with the distance to a would-be horizon. The ringdown frequencies and echo amplitudes depend on the perturbation closure and the matter near the barrier. By contrast, the leading echo delay is determined by the optical length.

\emph{Greybody cutoff.} The reflectivity of the barrier has an exponential greybody cutoff, $1-\Gamma\simeq e^{-(\omega-f)/T}$, with $f$ close to the real part of the least-damped prompt mode and $T$ about $10$--$15\%$ above the barrier-curvature value $\omega_I/\pi$. The cutoff is the same whether a horizon or a throat lies behind the barrier, and it does not depend on the surface gravity, so the prompt signal alone cannot tell the two apart. For the throat, however, the echoes return the transmitted part, a fraction $\Gamma$ of it in the first echo, and the whole signal has $|R_{\rm tot}|=1$ with no greybody cutoff. For a horizon the cutoff remains.

Several questions remain open: polar and fluid perturbations, which depend on the matter; the $\ell=0$ throat spectrum; the throat spectrum for $s\neq1$; and the backreaction and thermodynamics of quantum fields in the throat.

\section{Acknowledgments}

JA thanks Niayesh Afshordi, Babak Haghighat, Yun-Kau Lau, Puskar Mondal, and Hossein Yavartanoo for useful discussions, and the participants in Shing-Tung Yau's group meeting for helpful discussions.

\appendix
\section{Derivations}

\subsection{Master equation}
With $\sqrt{-g}=e^{\Phi}r^2\sin\theta/\sqrt F$ the wave equation is $e^{-\Phi}\sqrt F\,r^{-2}(e^{\Phi}r^2\sqrt F\,\psi')'+\omega^2e^{-2\Phi}\psi-\ell(\ell+1)\psi/r^2=0$. With $dr_*=e^{-\Phi}dr/\sqrt F$ and $\psi=\chi/r$ this gives Eq.~\eqref{eq:master}, using $d^2r/dr_*^2=e^{2\Phi}(F'/2+\Phi'F)$.

\subsection{Optical length}
With $u=\zeta\sqrt{\beta/\Delta}$, $\int d\zeta(\Delta+\beta \zeta^2)^{-(1+s)/2}=\Delta^{-s/2}\beta^{-1/2}\int du(1+u^2)^{-(1+s)/2}$, and the infinite-range integral is $B(\tfrac12,\tfrac s2)=\sqrt\pi\,\Gamma(s/2)/\Gamma(\tfrac{1+s}2)$. For $s=0$ it is $2\,\mathrm{arsinh}$, giving the logarithm of Eq.~\eqref{eq:twosided}.

\subsection{AdS$_2$ spectrum}
For $s=1$, with $F=\Delta\cosh^2\sigma$ and $N=N_0\sqrt\Delta\cosh\sigma$, $-N^2dt^2+d\rho^2=\ell_2^2(-\cosh^2\!\sigma\,d\tau^2+d\sigma^2)$ with $\ell_2=1/\sqrt\beta$ and $\tau=N_0\sqrt{\beta\Delta}\,t$. A scalar of mass $m^2=\ell(\ell+1)/R_M^2$ has normal modes $\omega_\tau=n+h$, $h(h-1)=m^2\ell_2^2$, in global time \cite{Higuchi2022}, hence $\omega_t=N_0\sqrt{\beta\Delta}(n+h)$. For axial gravitational waves $m^2=(\ell-1)(\ell+2)/R_M^2$, because $1-F\to1$ and $F'/2+\Phi'F\to0$ at the throat in Eq.~\eqref{eq:Vax}.

\subsection{Echo delay}\label{app:echo}
For $s>0$ and fixed $\omega$, the integral $\int\bigl[\omega/\sqrt{\omega^2-V}-1\bigr]dx$ over the cavity is split at the matching radius where $V\sim\omega^2$. Inside, $V\ll\omega^2$ and the integrand is $V/(2\omega^2)+O(V^2/\omega^4)$; with $dx=dr/(N\sqrt F)$ one has $V\,dx=N\,W\,dr/\sqrt F$, where $W=N^{-2}V$ is smooth and $N/\sqrt F=N_0F^{(s-1)/2}$. Near $r=R_M$, $F\simeq\Delta+\beta\zeta^2$ and $\int d\zeta\,(\Delta+\beta\zeta^2)^{(s-1)/2}$ converges as $\Delta\to0$ to $\int|\zeta|^{s-1}d\zeta$ for $s>0$. The remainder, from the region where $V\sim\omega^2$, is $\Delta$-independent because there $F=O(1)$ and $N=O(1)$.

\section{Curvature of the lapse family}\label{app:curv}

\begin{proposition}[curvature of $N=N_0F^{s/2}$]\label{prop:curv}
Let $F\simeq\Delta+\beta\zeta^2$ near the minimum and $N=N_0F^{s/2}$ with $s\ge0$.
\begin{enumerate}
\item In the static orthonormal frame (hats denote its components) the curvature components are $R_{\hat t\hat\rho\hat t\hat\rho}=N_{,\rho\rho}/N=s\beta+s(s-1)\beta^2\zeta^2/(\Delta+\beta\zeta^2)$, $R_{\hat t\hat\theta\hat t\hat\theta}=N_{,\rho}r_{,\rho}/(Nr)=s\beta\zeta/r$, $R_{\hat\rho\hat\theta\hat\rho\hat\theta}=-r_{,\rho\rho}/r=-\beta\zeta/r$ and $R_{\hat\theta\hat\phi\hat\theta\hat\phi}=(1-F)/r^2$. They are bounded uniformly in $\Delta\ge0$ for every $s\ge0$, so all curvature invariants are finite.
\item For $\Delta=0$ and $s>0$ the surface $\zeta=0$ is a degenerate Killing horizon ($N\to0$). In regular coordinates $(v,N)$, $v=t+r_*$, the metric is $-N^2dv^2+2c\,dv\,dN+r(N)^2d\Omega^2$ with $c=1/(s\sqrt\beta)$, the AdS$_2$ radius and $r-R_M\propto N^{1/s}$. The Riemann tensor is finite at the horizon if and only if $s=1$ or $0<s\le\tfrac12$. For $\tfrac12<s<1$ and $s>1$ the component $R_{N\theta N\theta}=-r\,r_{,NN}\propto N^{1/s-2}$ diverges, a parallelly propagated curvature singularity with finite scalar invariants. The metric is $C^\infty$ at the horizon only if $1/s$ is a positive integer.
\end{enumerate}
\end{proposition}
\emph{Proof sketch.} Part~1 follows from the warped-product formulas with $r_{,\rho}=\sqrt F$ and $N_{,\rho}/N=(s/2)F_{,\zeta}/\sqrt F$; the sectional curvatures reduce to Schwarzschild's for $N^2=F=1-2M/r$. For part~2 a symbolic computation of the Riemann tensor of the regular-coordinate metric gives $R_{N\theta N\theta}=-r\,r_{,NN}$, $R_{v\theta N\theta}=-Nr\,r_{,N}/c$, $R_{v\theta v\theta}=N^3r\,r_{,N}/c^2$, $R_{vNvN}=1$ (the AdS$_2$ part is constant), and $R_{\theta\phi\theta\phi}=r^2\sin^2\!\theta\,(c^2-N^2r_{,N}^2)/c^2$. With $\zeta\propto N^{1/s}$, $r_{,NN}\propto N^{1/s-2}$ times $(1/s)(1/s-1)$, which is finite for $1/s\ge2$ and vanishes at leading order for $s=1$: there $N=N_0\sqrt F$ is analytic in $\zeta$ with $N'(0)\ne0$, so $r(N)$ is analytic, e.g.\ for $F=\beta\zeta^2+\gamma\zeta^3$ one finds $r_{,NN}=-\gamma/(\beta^2N_0^2)$, finite; for $s>1$ even $r_{,N}\propto N^{1/s-1}$ diverges. Equivalently, the Einstein equations give $8\pi(\rho_E+p_r)=(s-1)F'/r$, so an infalling observer measures $T_{\ell\ell}\propto(s-1)\zeta^{1-2s}$, which diverges unless $s=1$ or $s\le\tfrac12$. $\square$

For $\Delta>0$ the lapse is positive and smooth, so the spacetime has no horizon and the curvature is finite for all $s$. Part~2 concerns only the exactly critical geometry and its degenerate horizon.

\section{Matter content}\label{app:matter}

\paragraph*{Isotropic fluids do not reach the threshold.}
An isotropic perfect fluid, $p_t=p_r=p$, is one possible closure, but it cannot reach the threshold. Conservation gives $p'=-(\rho_E+p)\Phi'$ with $\Phi'=(m+4\pi r^3p)/(r^2F)$, so near the throat $\Phi'\simeq K/(\Delta+\beta\zeta^2)$ with $K=1/(2R_M)+4\pi R_Mp>0$. Across the throat $\rho_E+p$ would then change by a factor $\sim e^{-\pi K/\sqrt{\beta\Delta}}$, which is incompatible with the constraint value $\rho_E\to1/(8\pi R_M^2)$ together with $p\ge0$. Integrating the TOV equation inward from $p(R)=0$ confirms this for the Burkert profile: the pressure diverges before the throat is reached.

\paragraph*{Energy conditions for general $s$.}
For $s\ne1$ the radial combination $8\pi(\rho_E+p_r)=(s-1)F'/r$ changes sign at the throat, so the null energy condition fails on one side of it: outside the throat ($R_M<r<r_c$) for $s<1$, and inside it ($r<R_M$) for $s>1$. Evaluating the stresses of Eq.~\eqref{eq:stresses} at $\Delta=10^{-3}$ out to $r_c$, we find that the null, weak and dominant conditions hold everywhere only for $s=1$, and for $s=1$ the strong condition fails only in the core, $r\lesssim0.15R$. For $s=0$ and $\tfrac12$ all four conditions fail outside the throat, $R_M<r<r_c$, where the null condition is violated, and for $s=\tfrac12$ the strong condition also fails in the core, $r\lesssim0.2R$. For $s=2$ all four fail inside the throat, $r<R_M$, and the dominant condition fails in addition for $0.35R\lesssim r\lesssim1.06R$, as the bound $s^2\beta R_M^2\le1$ ($s\le1.84$ for the Burkert profile) anticipates, and for $4.0R\lesssim r\le r_c$. At the centre of the throat, $\zeta=0$, the tangential pressure is $p_t=s\beta/(8\pi)$ for $\Delta>0$, and it approaches the AdS$_2$ value $s^2\beta/(8\pi)$ for $\sqrt{\Delta/\beta}\ll|\zeta|\ll R_M$. Of all the members of the family, $s=1$ therefore has the most physical matter content; the members with $s\ne1$ need effective matter that violates the null energy condition on one side of the throat.

\section{The cavity as a surface}\label{app:waveemb}

The embedding of Eq.~\eqref{z_general} gives a picture of the spatial slice. Figure~\ref{fig:waveemb} gives a picture of the axial waves instead. To construct it, consider a scalar wave on the two-dimensional surface
\begin{equation}
ds^2=dx^2+\rho(x)^2d\phi^2 ,\qquad x\equiv r_* .
\label{eq:waveemb}
\end{equation}
For $\psi=\rho^{-1/2}\chi(x)e^{im\phi-i\omega t}$ the wave equation $\psi_{,tt}=\nabla^2\psi$ gives
\begin{equation}
\chi_{,xx}+\Bigl[\omega^2-\frac{m^2}{\rho^2}-\frac{\rho''}{2\rho}+\frac{\rho'^2}{4\rho^2}\Bigr]\chi=0 ,
\end{equation}
where primes denote $d/dx$. If we choose $\rho=\sqrt{\ell(\ell+1)/V_{\rm ax}}$ and $m^2=\ell(\ell+1)$, the centrifugal term is exactly $V_{\rm ax}$, and the equation becomes Eq.~\eqref{eq:master} with the potential of Eq.~\eqref{eq:Vax}, up to curvature terms that vary on the scale of $\rho$ and drop out at high frequency. The surface thus turns the potential into geometry: where $V_{\rm ax}$ is small the surface is wide, and a high barrier becomes a narrow neck. Near the centre $V_{\rm ax}\simeq\ell(\ell+1)N_0^2/r^2$, so $\rho\simeq r/N_0$ and the surface closes. In the throat $V_{\rm ax}\simeq N^2(\ell-1)(\ell+2)/R_M^2$, so $\rho\propto1/N$ is large over the optical length $\Lambda$. At the barrier $\rho$ reaches its minimum, $\sqrt{\ell(\ell+1)/V_b}$, and far away $\rho\simeq r$. For a horizon, $V_{\rm ax}\propto F\to0$ and $\rho$ grows without bound. In the eikonal limit $\rho=r/N$, which is the circumference radius of the optical geometry $N^{-2}(dr^2/F+r^2d\Omega^2)$; its neck is the light ring.

A WKB wave of frequency $\omega$ propagates wherever $\rho>\sqrt{\ell(\ell+1)}/\omega$. For $\omega^2<V_b$ the neck is narrower than this and separates the bulb from the outside. A wave inside the bulb is trapped and leaks out by tunnelling once per round trip of length $2\Lambda$; this is the echo mechanism of Sec.~\ref{sec:echoes}. Above $V_b$ the wave passes through the neck, to the extent quantified by the greybody factor of Sec.~\ref{sec:greybody}.

Where $|\rho'|>1$, which happens on the walls of the bulb ($|\rho'|\le1.2$ for $s=1$, more for $s=2$), the surface~\eqref{eq:waveemb} cannot be embedded isometrically in flat space. Figure~\ref{fig:waveemb} therefore draws it with $x=r_*$ as the vertical coordinate, as in Fig.~\ref{fig:echo_pot}. Each horizontal circle has radius $\rho$, but vertical distances are optical lengths rather than lengths on the surface. The potential is computed on the evolution grid with the code of Sec.~\ref{sec:echoes}.

\section{Echoes of very long cavities}\label{app:fd}

The cost of an evolution grows as the square of the cavity length. For $s=2$ at $\Delta=10^{-5}$ and $10^{-6}$ the cavity is $2.2\times10^5$ and $2.2\times10^6$ long, which would require about $10^{14}$ and $10^{16}$ grid-point updates. For these cases we instead build the observed signal echo by echo from the terms of Eq.~\eqref{eq:Rtot}, the $n$-th echo being $T_b\bar T_b\bar R_b^{\,n-1}(\rho\,e^{2i\Phi})^n$ times the spectrum of the incident packet. The barrier amplitudes $R_b$, $T_b$, $\bar R_b$ and $\bar T_b$ are obtained by integrating Eq.~\eqref{eq:master} with a Numerov scheme from a matching point $x_b$ in the throat out to the observer, and the wall reflection $\rho$ by integrating from the regular centre out to a second matching point $x_a\le x_b$. Between the two, deep in the throat, $V_{\rm ax}$ is below $\epsilon=2\times10^{-6}$, far below $\omega^2$ for every frequency the barrier transmits, and it varies on scales much longer than $1/\omega$. There we propagate the wave by WKB, $\Phi(\omega)=\int_{x_a}^{x_b}\sqrt{\omega^2-V_{\rm ax}}\,dx$. When $V_{\rm ax}$ never drops below $\epsilon$, as in all evolved cases except $s=2$ at $\Delta=10^{-4}$, we take $x_a=x_b$ at the minimum of the potential, and the factorization is exact. Each echo is synthesized on a short time window after the linear phases $\omega(x_b-x_a)$ and the block lengths have been taken out, and its arrival time is read off as the peak of the same smoothed envelope used for the evolutions. Frequencies below $0.15$, which the barrier does not transmit, are tapered off.

Where both methods can be used ($s=\tfrac12$ and $1$ at $10^{-6}\le\Delta\le10^{-3}$, $s=2$ at $10^{-4}\le\Delta\le10^{-3}$), the echo spacings agree with the evolutions to within $0.1\%$ (within $2\%$ for the short cavity of $s=\tfrac12$ at $\Delta=10^{-3}$), and the echo amplitudes to within $5\%$. For $s=2$ at $\Delta=10^{-4}$, changing $\epsilon$ tenfold or halving the Numerov step changes the spacings by less than $0.01$, that is, by less than $10^{-4}\%$. The same computation provides the $s=2$ points at $\Delta=10^{-5}$ and $10^{-6}$ in Fig.~\ref{fig:echo_scal} and follows the echo trains out to $40$ echoes.

\bibliography{paper_merged_v25}

\end{document}